\documentclass[12pt]{article}
\usepackage{amsmath}
\usepackage{graphicx,psfrag,epsf}
\usepackage{enumerate}
\usepackage{natbib}
\usepackage{url} 
\newcommand{\vn}[1]{\vnform{#1}}
\newcommand{\vnform}[1]{\mathtt{#1}}

\newcommand{\blind}{0}

\begin{document}

\def\spacingset#1{\renewcommand{\baselinestretch}%
{#1}\small\normalsize} \spacingset{1}


\if0\blind
{
  \title{\bf Using Tableau and Google Map API for Understanding the Impact of Walkability on Dublin City}
  \author{Minkun Kim \\
    School of Computing, Dublin City University\\
    email: minkun.kim4@mail.dcu.ie \\
    }
  \maketitle
} \fi

\if1\blind
{
  \bigskip
  \bigskip
  \bigskip
  \begin{center}
    {\LARGE\bf Using Tableau and Google Map API for Understanding the Impact of Walkability on Dublin City}
\end{center}
  \medskip
} \fi

\bigskip
\begin{abstract}
In this article, we explore two effective means to communicate the concept of walkability - 1) visualization, and 2) descriptive statistics. We introduce the concept of walkability as measuring the quality of an urban space based on the distance needed to walk from that space to a range of different social, environmental, and economic amenities. We use Dublin city as a worked example and explore quantification and visualization of walkability of various areas of the city. We utilize the Google Map API and Tableau to visualize the less walkable areas across Dublin city and using WLS regression, we assess the effects of unwalkability on house prices in Dublin, thus quantifying the importance of walkable areas from an economic perspective.
\end{abstract}

\noindent%
{\it Keywords:}  Google Map API, Tableau, Walkability, Dublin City, Quality of Life, Date driven Urbanism, House price prediction
\vfill

\newpage
\spacingset{1.45} 
\section{Introduction}
\label{sec:intro}
Cities around the world aspiring to draw a talented young workforce start focusing on creating a good pedestrian environment. A recent study by \cite{rafiemanzelat2017city} has shown that walkable neighborhoods contribute to the improvement of Quality of Life in terms of health, social life and economic growth. However, understanding walkability in a city is not always simple due to its difficulty with visualization, and its complex association with the urban economy. This paper explores visualization and descriptive statistics to assess walkability along the pedestrian street network in Dublin and to understand its impact on the local economy.
\begin{figure}[!ht] 
    \begin{center} 
        \includegraphics[width=1\textwidth]{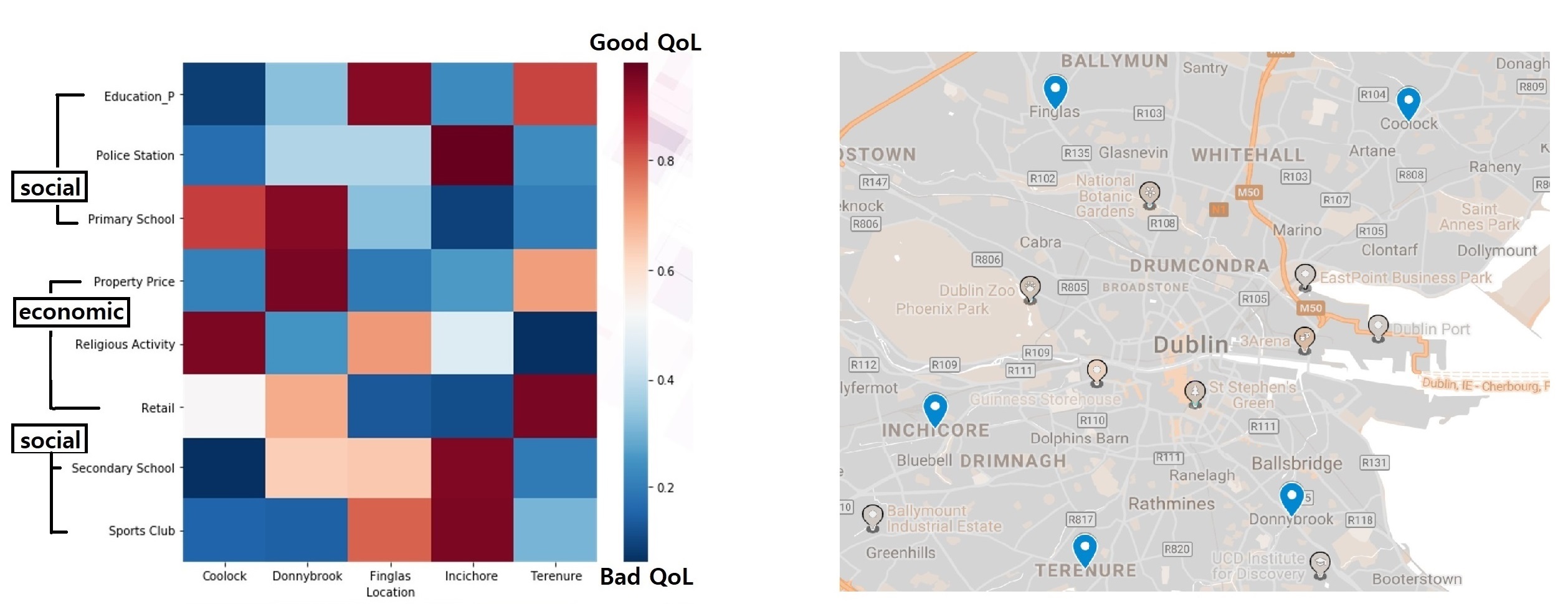}
    \end{center}
    \caption{Quality of Life Scoring Matrix produced by the Insight Centre team at DCU (2018). It compares the Quality of Life in the five random areas in Dublin - Coolock, Donnybrook, Finglas, Inchicore, Terenure}
\end{figure}
Commissioned by Dublin City Council, the research team in the Insight Center for Data Analytics at DCU - Prof.Alan Smeaton, Mr.George Mihailescu, Mr.Lavleen Bhat, and Dr.Kevin McGuinness (2018) - explored some methods to understand people's life quality in Dublin by introducing computer science approaches in the traditional urban planning process. Although their work was not published, they developed the analysis pipeline generating the ``Quality of Life index" by computing the expected quality of life at any given coordinate in Dublin. Its output is the scoring matrix shown in Figure (1) based on economic, social, and environmental indicators such as amenity proximity, property price, air quality, etc. This data-driven approach to site analysis can speed up the traditional urban planning process. For example, when deciding new development target area, urban planners would be able to make a data-driven comparison in the ``Quality of Life index" between any different candidate locations and hence, would be able to evaluate the performance of their new urban interventions.  

However, simply relying on the exact location of a certain area, their approach is not able to explain the entire area of Dublin. In addition to this, their scoring matrix for the entire area causes confusion due to the inconsistent color-coded values. In spite of such issues, their work still offers useful insight as its proximity analysis exemplifies an effective \textbf{walkability} measurement method, reflecting spatial differences explained by the data. In this paper, we would like to focus on the \textbf{walkability} measurement.

From the perspective of urban economy, there are interesting studies that praise the walkable neighborhood by assessing the impact of walkability on the city's economic success. For example, \cite{glaeser2011triumph} points out that the success of Silicon Valley is largely based on its fantastic walkable environment coupled with the great weather of San Francisco. He does not mean that the walkability of the city is the direct reason for the city's economic success, but its walkability facilitates its success by giving more chances to socialize and interact with talented young people who eventually bring success. And the success is in line with the improvement of the city's quality of life in the long run. Some researchers try to understand walkability in relation to traveler's socioeconomic status, using mobile phone data. \cite{li2018investigating} at MIT Senseable City Lab developed six walking activity indicators in order to quantify important characteristics of walking activity and compare their statistical properties. This is followed by the analysis of socioeconomic classes by \cite{xu2018human}, asking questions whether the rich tend to spend more time in walkable areas or not. The result of the study shows that the relationship between walking behavior and traveler's socioeconomic status could vary across cities because it is influenced by the spatial arrangement of the activity locations (or points of interest) in each city. Another strand of walkability research can be conducted in relation to the distance from the city center and real estate values. \cite{d2019quality} quantifies the relationship between characteristics of an area and its property values by using an OLS regression where the quality of the area is \(X_{1}\), the distance to the city center is \(X_{2}\), and the real estate value is \(Y\). This research addresses walkability as one of the important qualities impacting real estate values and delineates how the walkable environment can bring superior economic returns to countries, regions, firms, and individuals. 

\section{Data Gathered}
\noindent
\textbf{Data-01: For Visualising Walkability:} We investigate the walkability of Dublin through the footpath calculation, exploring all possible pedestrian routes from each origin (amenity point) to every destination (random grid point) located within a radius of 2Km from the origin. This task is conducted on the two datasets shown in Figure (2) generously provided by the Dublin City Council. The details are as follows.
\begin{itemize} 
    \item \textbf{DCC-sap.KML}: Provided by Dublin City Council, it contains coordinates of all address points available in Dublin City Area. We use them as the possible \underline{destination points} in the footpath calculation. The total record size is 246,617.    
    \item \textbf{NACE-codes.csv}: Provided by Dublin City Council, it contains coordinates and other related information of all amenity places in Dublin. If a place is not involved in any business activities, NACE dataset does not provide any information on the place. We collected the coordinates and some information on 1) Primary schools, 2) Secondary schools, 3) Universities/Institutes, 4) Religious Activities, 5) Medical services, 6) Retail, shops/restaurants, 7) Sports clubs, and 8) Garda stations in Dublin. We use them as the \underline{origin points} in the footpath calculation. The total record sizes are  different by each section:
    \begin{itemize}
        \item 1) Primary schools: 216 
        \item 2) Secondary schools: 65 
        \item 3) University/Institutes: 49 
        \item 4) Religious Activities: 368
        \item 5) Medical services: 600 
        \item 6) Retail: shops/restaurants: 405 
        \item 7) Sports clubs: 170 
        \item 8) Garda stations: 60 
    \end{itemize}
\end{itemize}
\begin{figure}[!ht] 
    \begin{center} 
        \includegraphics[width=1\textwidth ]{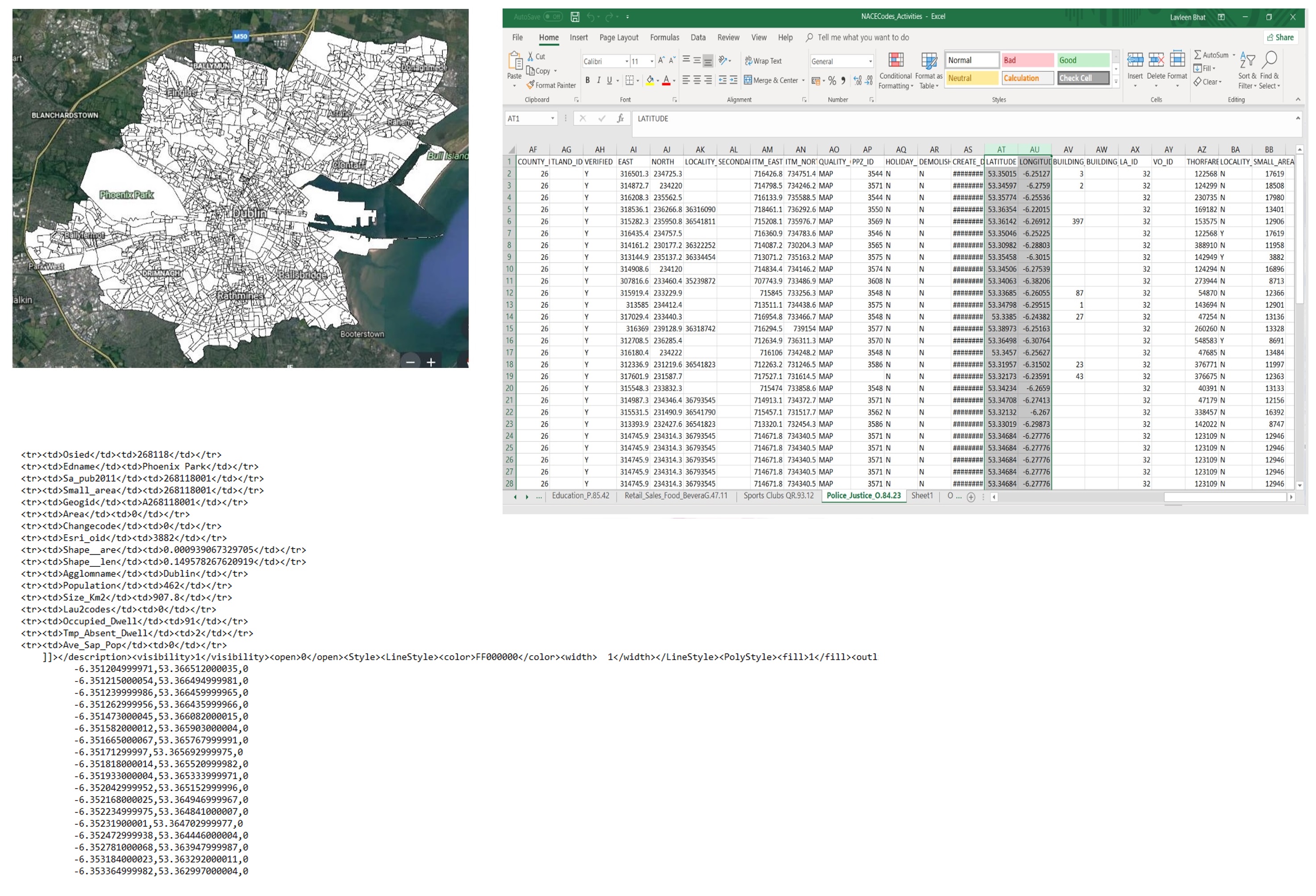}
    \end{center}
    \caption{\textbf{Left:} KML file loaded into Google Earth. The figure illustrates the coordinate points of all areas accessible for this project; \textbf{Right:} NACE Codes dataset. It groups organizations according to their business activities and provides relevant information with consistent items}
\end{figure} 

\noindent
NACE stands for``Nomenclature statistique des Activités économiques dans la Communauté Européenne" in French.
NACE codes have the purpose of developing the system to give a common basis for statistical classifications of economic activities. It helps develop the comparability of statistics produced in different domains associated with any business activities. The code is subdivided into a hierarchical, four-level structure - section, division, group, and class. Once we choose the correct section, we can get a dataset that includes other items such as division, group, and class. NACE code gives a series of classified datasets according to our query in the section. Our dataset is obtained from the sections of \(O/QR/G/P/S\). For example, we can use the Section code: 1) \textit{P85.20} for Primary schools; 2) \textit{P85.30} for Secondary schools; 3) \textit{P85.42} for University/Institutes; 4) \textit{Q.86.90} for Medical services; 5) \textit{G47.11} for Retail-shops/restaurants; 6) \textit{QR93.12} for Sports clubs; 7) \textit{O84.23} for Garda stations; 8) \textit{S94.91} for Religious Activities. Each section brings multiple items such as division, group, class, building ID, address, usage, construction information, coordinates, commercial delivery points, etc. In order to investigate the walkability of Dublin, we aim to understand the differences in the total travel distance between the Euclidean method and Google Maps' shortest footpath method. As for the \underline{origin points}, we already have the ``NACE codes" dataset in CSV format that describes every amenity point available in Dublin. \\
\begin{figure}[!ht] 
    \begin{center} 
        \includegraphics[width=1\textwidth]{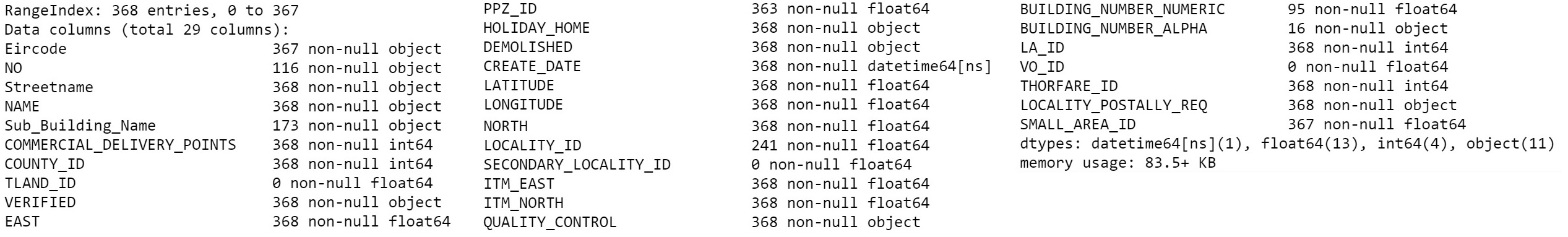}
    \end{center}
    \caption{Consistent items from the NACE-codes dataset}
\end{figure} 

We select \textbf{Name, Latitude, Longitude, Commercial-Delivery-Points} that are enough information to describe each origin (amenity) point. As for the possible \underline{destination points}, parsing the ``DCC-SAP" file in KML format with a Python library (``beautifulsoup4") gives all coordinates available that represent the possible destination areas over Dublin. They are random grid points, thus the location information is all we need.\\

\noindent
\textbf{Data-02: For Understanding Walkability:} We investigate if walkability, as one of the measures for the Quality of Life, is a decisive factor for the formation of real estate values. A simple statistical analysis can be conducted on the dataset collected from DAFT, an online property marketplace: \textit(www.daft.ie). 
\begin{itemize}
    \item \textbf{Daft-House-Prices-2017.csv:} It contains coordinates, address, type of dwelling, and prices of all housing properties available in Dublin in 2017. 
\end{itemize}

\section{visualization steps}
Given that we obtain all coordinates of places available in Dublin and our first goal is to visualize the \textbf{differences in the travel lengths} calculated by Euclidean method and Google Map algorithm, our visualization procedure involves 5 steps - 1. Generate 1000 random destinations all over Dublin, 2. Define the \textbf{importance weights} and assign them to each amenity point, 3. Compute straight lines with the Euclidean method, 4. Generate shortest paths with Google Map algorithm to compare with the straight lines, and 5. Visualize all the differences and the footpaths that contribute to these differences.\\ 

\noindent
\textbf{Step 01. Pre-processing: 1000 Random Destinations}: Using the \textit{BeautifulSoup4} Python library, we could pull all addresses and coordinates available in Dublin from the KML file. The total record size is 246,617. Plotting all the coordinates helps us determine the boundary of values and generate 1000 random points arranged on a grid. For example, once we get the maximum and minimum value of the coordinates, then we can use the Numpy library with the function: \textit{'np.arange(min-coordinate, max-coordinate, stepsize)'} to generate all coordinates of random destinations in Figure (4). \\
\begin{figure}[!ht] 
    \begin{center} 
        \includegraphics[width=0.8\textwidth]{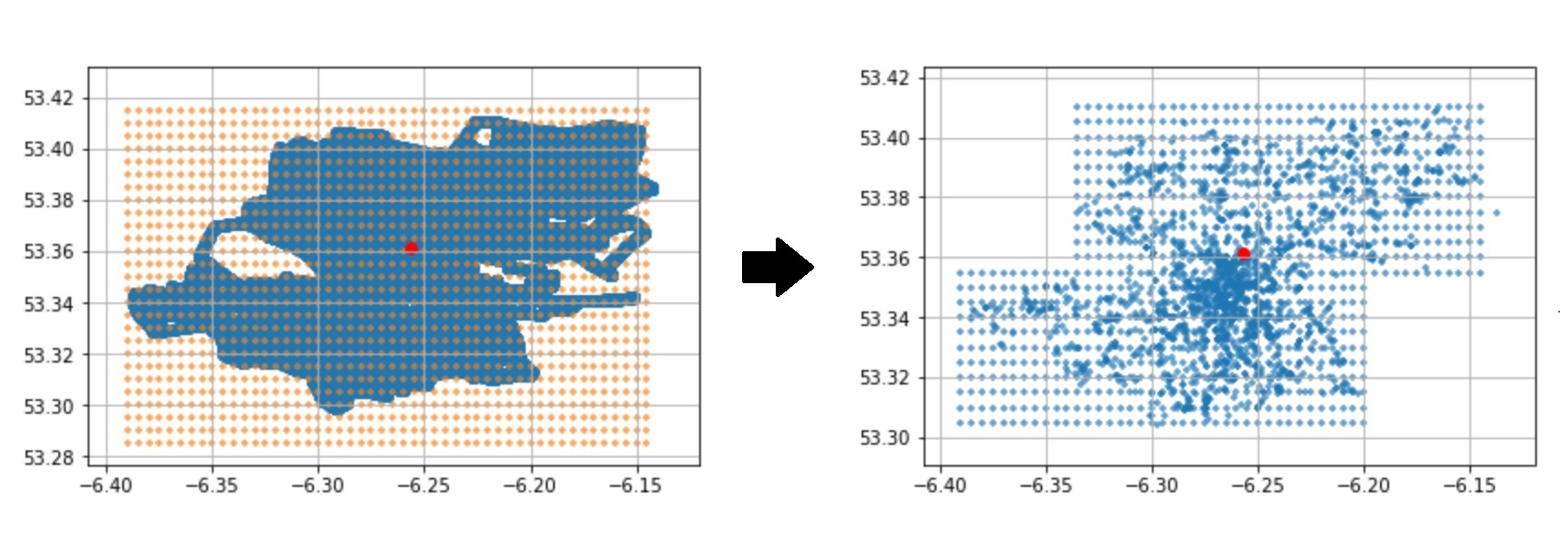}
    \end{center}
    \caption{Destinations: Generating 1000 Random Grid Points in Step 01 and their adjustment on the Dublin's surface, produced with \textit{matplotlib}. The origins and destinations are shown together in the plot.}
\end{figure} 

\noindent
\textbf{Step 02. Pre-processing: 1900 Amenity Origins}: From the original NACE-codes dataset, we obtain - 1) property names, 2) coordinates, and 3) a number of commercial-delivery points and they are enough information to understand each amenity point. In particular, we utilize the ``number of commercial-delivery points" to effectively differentiate each point. If a certain amenity point holds more commercial delivery points, one can see that more logistics are involved, and this possibly indicates more population, and more social, and economic activities are expected on the point. It is obvious that walkability would be different according to the importance of the place because of the extent of pedestrian traffic. Therefore, we define the importance weight for each amenity point by reflecting the individual size of commercial deliveries associated with these points shown in Figure (5). The size of the minimum delivery point is 1 and that of the maximum delivery point is 17 which is held by Ilac Shopping Mall, for example.\\
\begin{figure}[!ht] 
    \begin{center} 
        \includegraphics[width=1\textwidth]{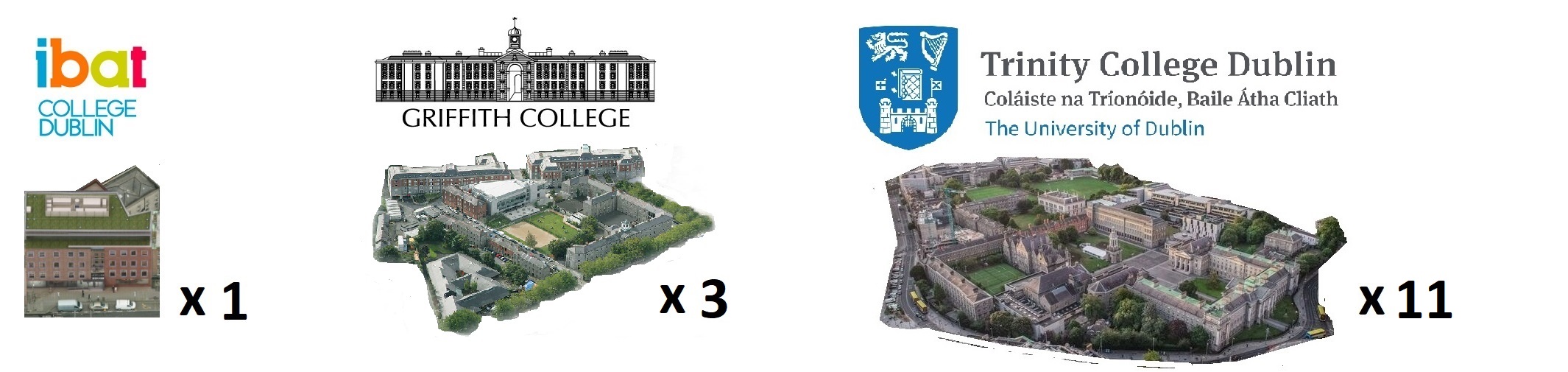}
    \end{center}
    \caption{What is the Importance weight?; We use the ``number of commercial delivery points". For example, in Step 02, although they are all tertiary education providers, Trinity College, Griffith College, and IBAT College are different in terms of size of activities, population, logistics, etc.}
\end{figure} 
 
\noindent
\textbf{Step 03. Computing straight lines with Euclidean distance}: From the pre-processing step, we can obtain 1000 grid points as destinations and 1900 amenity points as origins. Now we compute the shortest travels between each origin and destination pair in an Euclidean sense. However, given that using all grid points to measure the proximity of each amenity point is computationally expensive, we can filter out some grid points that do not fit within the boundary of the walkability circle for each calculation. That is to say, if more than 2 km distance, in general, cannot be considered walkable, we only take the grid points that are within the radius of 2 km from each amenity point.
\noindent
Since we have the coordinates of each point, and all coordinates imply the locations on a sphere, we compute the Euclidean distances on a sphere between the origins and the filtered destinations, using the \textit{Haversine Formula} suggested by \cite{EuclideanDistance}. This is expressed in trigonometric function: \( haversine(\theta) = \sin^{2}(\theta/2) \) as shown below.\\

\( dist[(x,y), (a,b)] = \sqrt{(x-a)^2 + (y-b)^2} \) is \\
$$2R\sin^{-1}\biggl(\sqrt{\sin^{2}(\frac{\phi_{2}-\phi_{1}}{2})+\cos(\phi_{1})\cos(\phi_{2})\sin^{2}(\frac{\lambda_{2}-\lambda_{1}}{2})}\biggr)$$ \\
where R is the radius of earth(6371 km), \(dist[(x,y), (a,b)]\) is the distance between two points, \(\phi_{1},\phi_{2}\) is latitude of the two points, and \(\lambda_{1},\lambda_{2}\) is longitude of the two points respectively. After collecting the calculations that return the values greater than 1.5 km and less than 2 km, we get the data frame with the size of 4862 as below. \\
\begin{figure}[!ht] 
    \begin{center} 
        \includegraphics[width=1\textwidth]{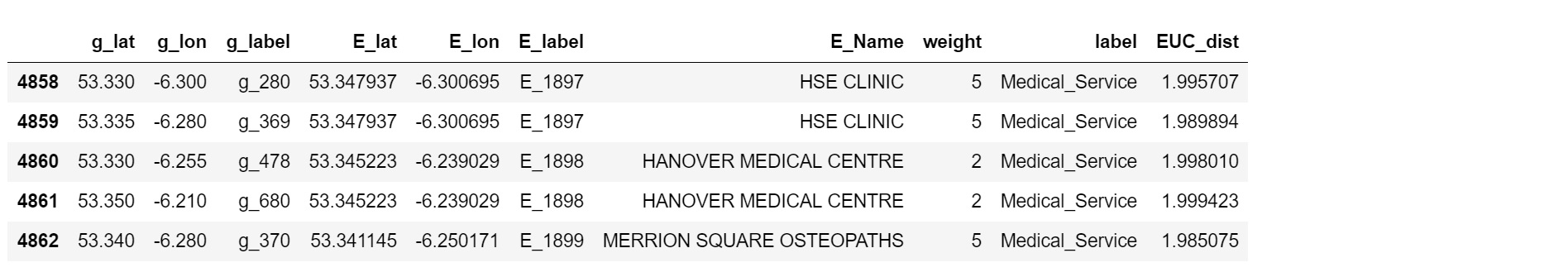}
    \end{center}
    \caption{Pandas Dataframe with the last 5 records of Euclidean distance calculations between the origin and the selected destinations, using Haversine formula; 4863 calculations in total}
\end{figure}

\noindent
\textbf{Step 04. Computing footpaths with Google Map API}
: Using an HTTP request, we can have Google Map Directions API calculate the most efficient footpath between an origin coordinate to a destination coordinate. When building the HTTP request, and setting up some parameters (mode, metric, etc), we can get JSON data containing each foot travel length, and the waypoints that make these travels accordingly. This API runs one calculation at a time, thus we make 4862 runs in total. 
\begin{itemize}
    \item{API Request and JSON parsing:}
By setting up parameters such as `metric' as a measuring unit and `walking' as a travel mode, we can obtain high-quality JSON data from the API Request. The total foot-travel length can be collected from the first part of the `legs' in the JSON data and the series of waypoints for each travel can be collected from the 'steps' in the JSON data. Tableau path plot requires the order label and coordinate of each waypoint. Hence, we encode all the necessary information with string data type and add them to the new data frame for this plot. Feeding the coordinates of each waypoint and their respective labels that order these waypoints into Tableau, we can plot all footpaths calculated by Google Map API. Once visualizing all footpaths, we can understand the walkability problem more efficiently by the corresponding area as each footpath illustrates the problematic travels in detail.
\end{itemize} 
\begin{figure}[!ht] 
    \begin{center} 
        \includegraphics[width=1\textwidth]{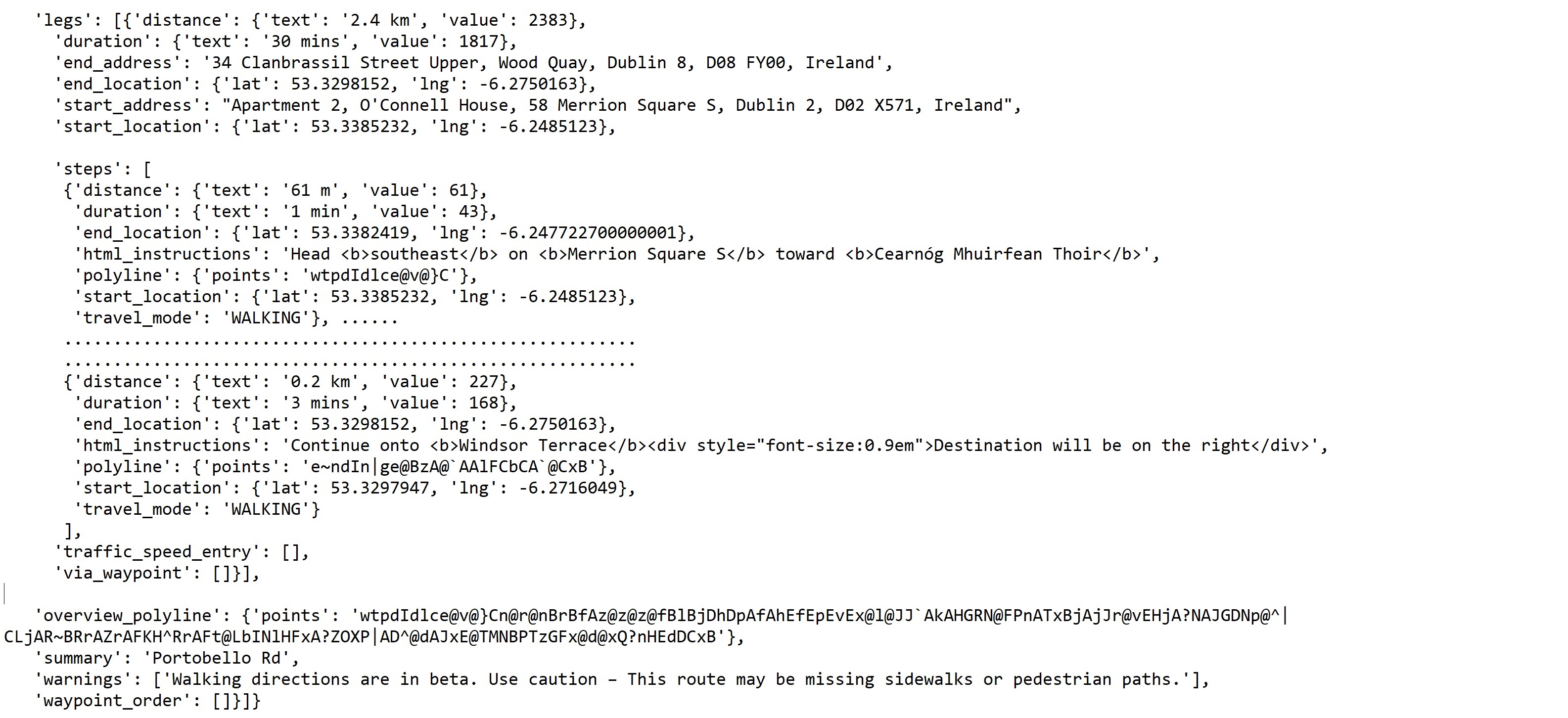}
    \end{center}
    \caption{API request output in JSON. The first part of the `legs' gives summary information, and `steps' gives each individual information on waypoints that make the travel.}
\end{figure} 
\noindent 
We collect the 5 problematic travel routes (origin-destination pairs) per each amenity point (origin) that demonstrate the biggest discrepancy in total travel length. This helps us to capture the most problematic areas in terms of walkability. This subset is utilized to create the visualization of the discrepancies in Tableau.
\begin{figure}[!ht] 
    \begin{center} 
        \includegraphics[width=1\textwidth]{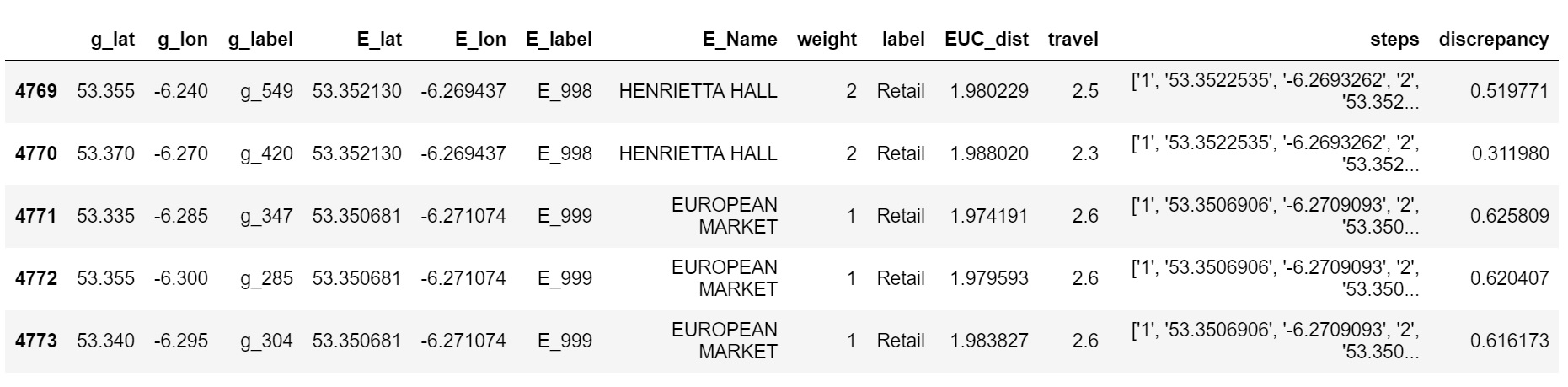}
    \end{center}
    \caption{Pandas data frame with the last 5 records that demonstrate the biggest discrepancy in the total travel length between the footpath and the Euclidean perception and the encoded waypoints (shown in ``steps" column) collected from Google Map API.}
\end{figure} 

Lastly, in order to create the footpaths visualization in Tableau, it is necessary to parse the waypoint information encoded in the string shown in the ``steps" column of the data frame above. Each waypoint consists of a series of order labels and coordinates, thus we can create a new data frame by converting the strings into six lists - PointID, Latitude, Longitude, weight, label, and travel length. \\    

\begin{figure}[!ht] 
    \begin{center} 
        \includegraphics[width=1\textwidth]{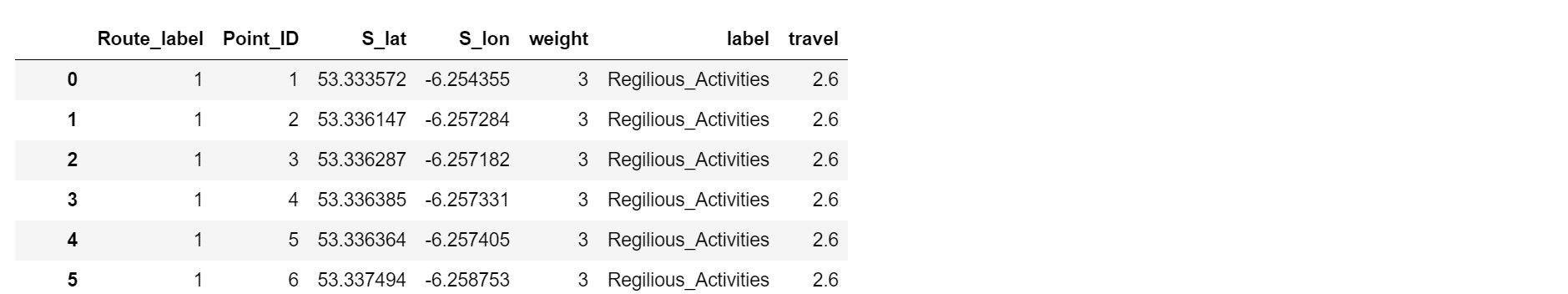}
    \end{center}
    \caption{Pandas data frame that shows labels and coordinates to visualize footpath in Tableau; Total size:53,627}
\end{figure}

\noindent
\textbf{Step 05. Feeding data into Tableau} 
: In Tableau, it is important to make clear what visual components constitute the visualization and which component each data point belongs to. Having three major components - origins, circles, and destinations -, we can differentiate each data point by labeling 1)point, 2)circle, and 3) random grid.\\

\noindent
The ``random grid" refers to the 1000 destination coordinates. The ``point" gives the information on each origin (amenity point) - coordinates, facility names, number of commercial delivery points. Generated from the amenity point, the ``circle" is a bunch of pseudo points encircling each amenity point by specifying degrees - 0, 15, 30,..360. We can have these circles react to the modification in their sizes and numbers by parameterizing the two variables - \textbf{``importance weights"} and \textbf{``differences in the travel lengths"} in the dataset. This can be done by the ``calculated field" expression in Tableau as follows:

- Generating Latitudes for the circles: \textbf{Lat~Circle} \\
\noindent  
\( IF [\textbf{label}]=``Circle" THEN \\
DEGREES( ASIN(SIN(RADIANS([\textbf{Latitude}]))*\\
COS(([\textbf{scale}]*[\textbf{differences in the travel}])/6371)+\\ COS(RADIANS([\textbf{Latitude}]))*\\
SIN(([\textbf{scale}]*[\textbf{differences in the travel}])/6371)*\\
COS(RADIANS([degree]))) )\\
ELSE [\textbf{Latitude}]\\
END \)

- Generating Longitudes for the circles: \textbf{Lon~Circle} \\
\noindent  
\( IF [\textbf{label}]=``Circle" THEN \\
DEGREES( RADIANS([\textbf{Latitude}]) + \\ ATAN2(COS(([\textbf{scale}]*[\textbf{differences in the travel}])/6371) - \\
SIN(RADIANS( [\textbf{Latitude}] ))* \\
SIN(RADIANS( [Lat~Circle] )), \\
SIN(RADIANS([degree]))* \\
SIN(([\textbf{scale}]*[\textbf{differences in the travel}])/6371)* \\
COS(RADIANS( [\textbf{Latitude}] ))) )-90 \\
ELSE [\textbf{Longitude}] \\
END \)
\begin{figure}[!ht] 
    \begin{center} 
        \includegraphics[width=1\textwidth]{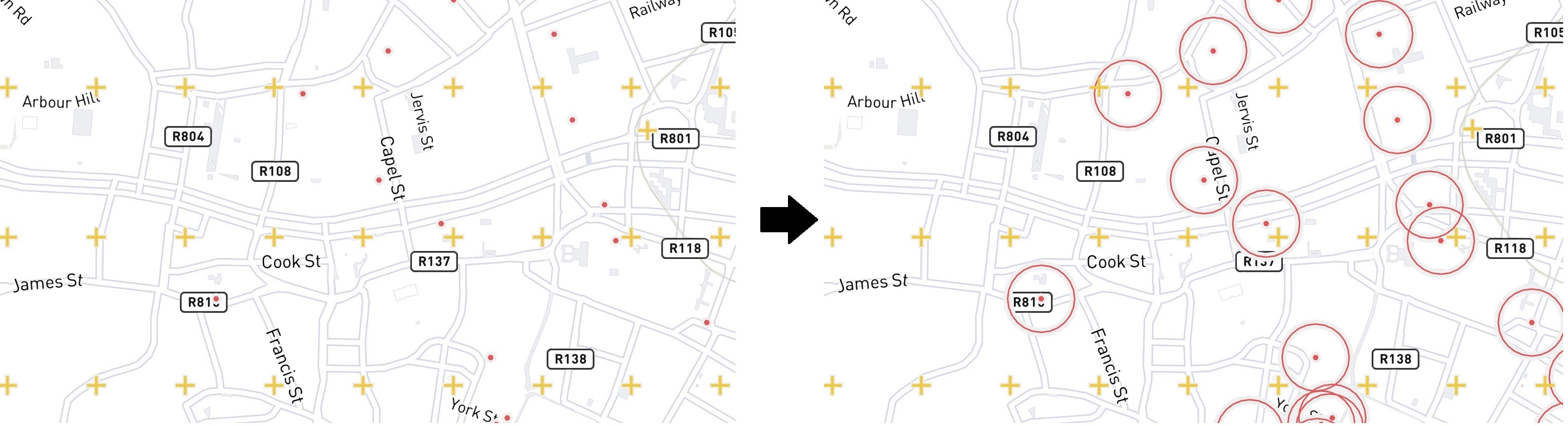}
    \end{center}
    \caption{Example: Generating Circles in Tableau; The red points are origins (amenity points) and the yellow points are destinations; Circles are generated from the origins. The size and number of circles are adjustable by parameterizing the variables.}
\end{figure}

\section{Mapping Walkability: Visualization}
Our walkability visualization does not depend solely upon a single feature or a single plot. There are lots of different features such as circle size, importance weight, activity label, property name, Euclidean value, foot travel value, foot travel waypoint, etc that are required to be added, dropped, or mixed depending on situations to produce meaning. If we want to understand the detailed conditions by investigating the interplay of such features, Tableau visualization can be an excellent tool to tackle such issues thanks to its interactive display. We exploit Tableau's interactive functionality in three different ways.
\begin{itemize}
    \item Using Circle Size referring to the \textbf{difference in the importance weight} between amenity points.
    \item Using Circle Size referring to the \textbf{difference in the travel lengths} calculated by Euclidean method and Google Map’s algorithm; Plus coupling the plot with the importance weights. 
    \item Using Circle size referring to the \textbf{difference in the travel lengths} calculated by Euclidean method and Google Map’s algorithm; Plus coupling the plot with the footpaths articulation.\\
\end{itemize}

\noindent
\textbf{1. Importance Weight Plot}: This visualization uses Circle Size to deliver the \textbf{difference in importance weight} between amenity points. The main goal of this plot is to better understand the amenity points and get an idea of how to address them. Each amenity point comes with four attributes - Name, Coordinates, Importance weight, and activity label. Using the weight information, Tableau generates a series of circles different in size by the activity label. The bigger circles indicate activities on a bigger scale and thus hold a greater importance.\\
\begin{figure}[!ht] 
    \begin{center} 
        \includegraphics[width=1\textwidth]{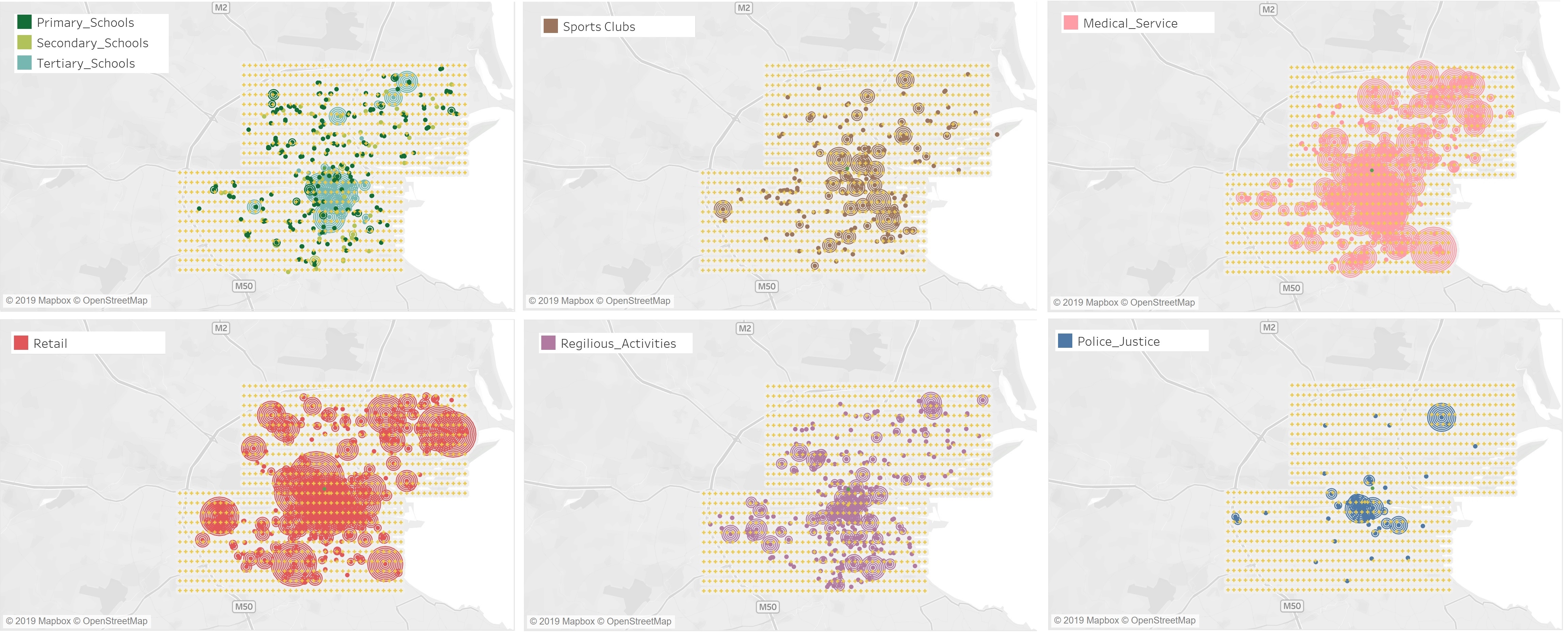}
    \end{center}
    \caption{\textbf{Importance Weight Plots:} Education (Tertiary, Secondary, Primary), Sports Club, Retail, Religious Facility, Garda office, Medical Service}
\end{figure}

\noindent
From the visualization above, there are notable differences in the distributions of the importance weights between different activities. For example, there is not much variation in the importance between Primary and secondary schools, and a medium variation is observed between Higher institutions, Sports Clubs, Religious places, and Garda Offices, while Retail and Medical Services register huge differences in the importance by individuals. This should be reflected when we examine foot traffic from each amenity point as the amenity point with bigger importance obviously triggers more foot traffic.  \\         

\noindent
\textbf{2. Foot-path Discrepancy Significance Plot}: This visualization below uses the Circle Size representing the \textbf{difference in the footpath length} calculated by the Euclidean method and Google Map algorithm, then coupling the plot with the importance weights. We discussed the meaning of the importance of weight previously. Coupling the concept of a foot-path discrepancy between the perception and the reality with the importance weight, one can see if the foot-path discrepancy discovered is significant or not. 
\begin{figure}[!ht] 
    \begin{center} 
        \includegraphics[width=1\textwidth]{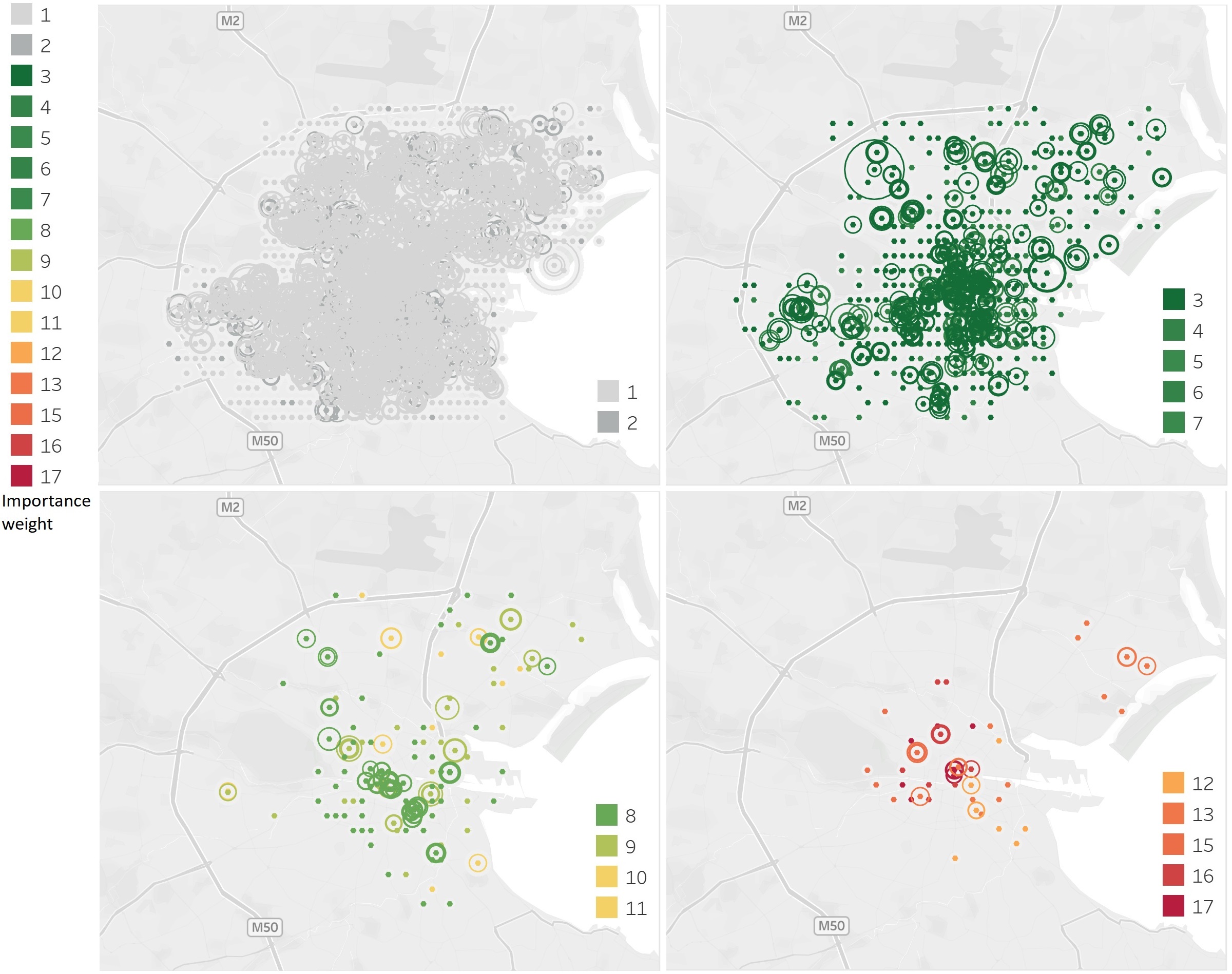}
    \end{center}
    \caption{\textbf{Foot-path Discrepancy Significance Plot:} All activities are included. The bigger circle indicates the area is less walkable because of the bigger discrepancy. The stronger color indicates the bigger importance of the amenity point.}
\end{figure}
It is interesting to note that the discrepancies between the activities with higher importance weight(red colored) are likely to be less significant while the majority of greater significance in the discrepancy is observed between the amenity points with less importance. \\  

\begin{figure}[!ht] 
    \begin{center} 
        \includegraphics[width=0.6\textwidth]{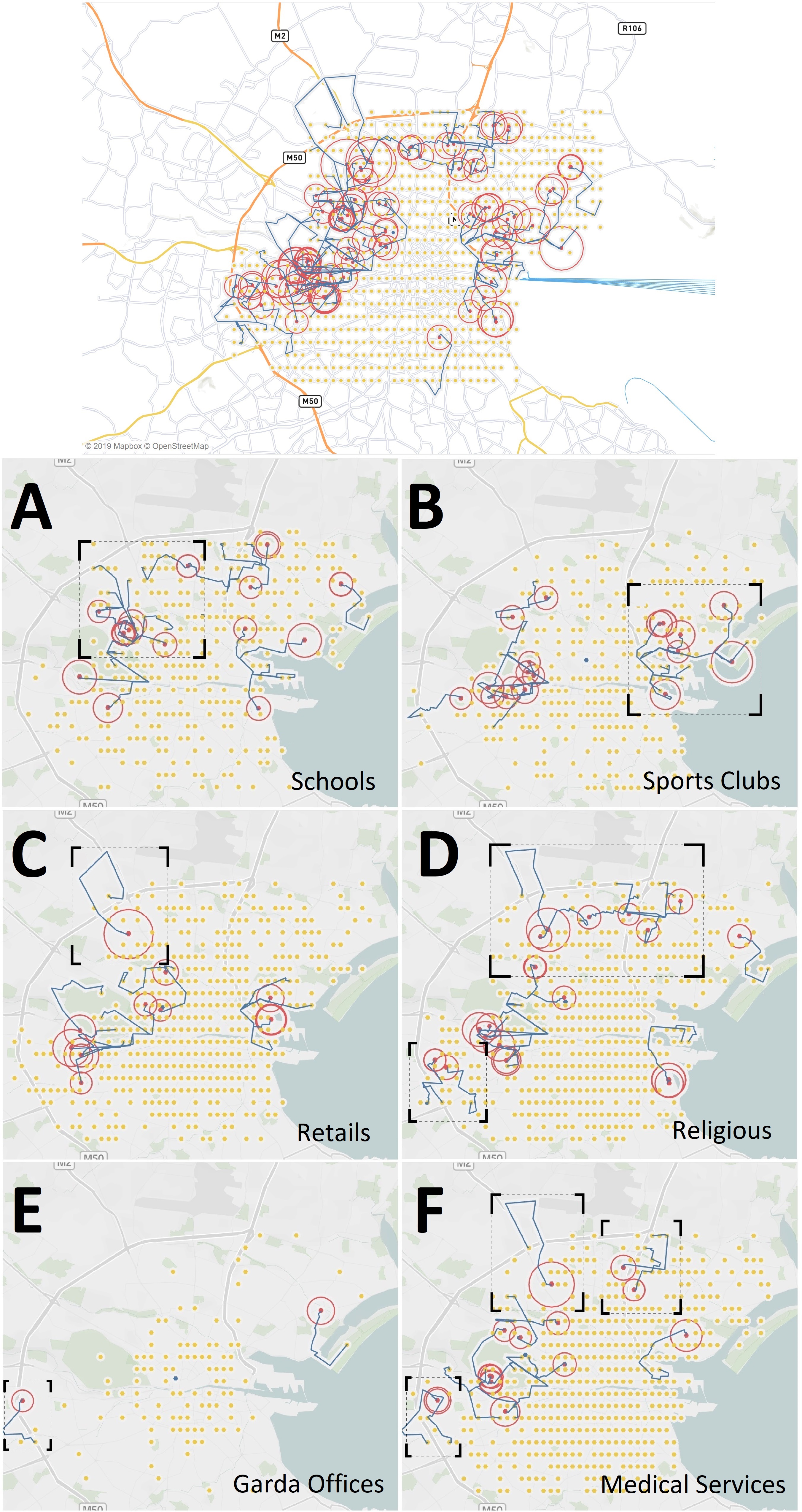}
    \end{center}
    \caption{Top. Foot-Path discrepancy Plot with the Articulation of the Foot-paths; Below. Sampling the most problematic areas \textbf{A,B,C,D,E,F} and footpaths} 
\end{figure}
\noindent
\textbf{3. FootPath discrepancy Plot with  Articulation of the Footpaths}: We now investigate walkability in Dublin in a more comprehensive manner. Figure (13) reveals all problematic amenity points and corresponding problematic footpaths to the nearest origin that take more than 4-9 km from the origin to the destination even though both are located within a radius of 2 km. Interestingly, the majority of the problematic areas are likely to appear around the seashore or the North and West ends of Dublin, but why? If that is the case, in Tableau, we can zoom in on the areas of our interest so that we can examine more deeply and more details come into view. 

Here, we pose a question: \textbf{What makes the areas unwalkable?} With regard to this question, the researchers in Larry Seeman \cite{MeasuringWalkability} proposed four measures to evaluate walkability as follows:
\begin{itemize} 
    \item \textbf{Continuity}: Is the network free from gaps and barriers?
    \item \textbf{Street Crossings}: Can the pedestrian safely cross streets?
    \item \textbf{Visual Interest}: Is the environment attractive and comfortable?
    \item \textbf{Security}: Is the environment secure and well-lighted with a good line of sight to see the pedestrian? 
\end{itemize}
One can see that these measures above can be applied consistently throughout the sample areas described as follows.\\

\noindent
\textbf{A. Footpath from Schools (Primary/Secondary/Tertiary)}: In Figure (14), the urban infrastructure here unfortunately discourages pedestrian street crossing because of the emphasis on vehicle traffic flow. This area lacks a grid system and its curvilinear street patterns add additional distance to the footpaths. Multiple primary schools here are located above Phoenix Park which creates a huge green island in the middle of the city. \\
\begin{figure}[!ht] 
    \begin{center} 
        \includegraphics[width=0.9\textwidth]{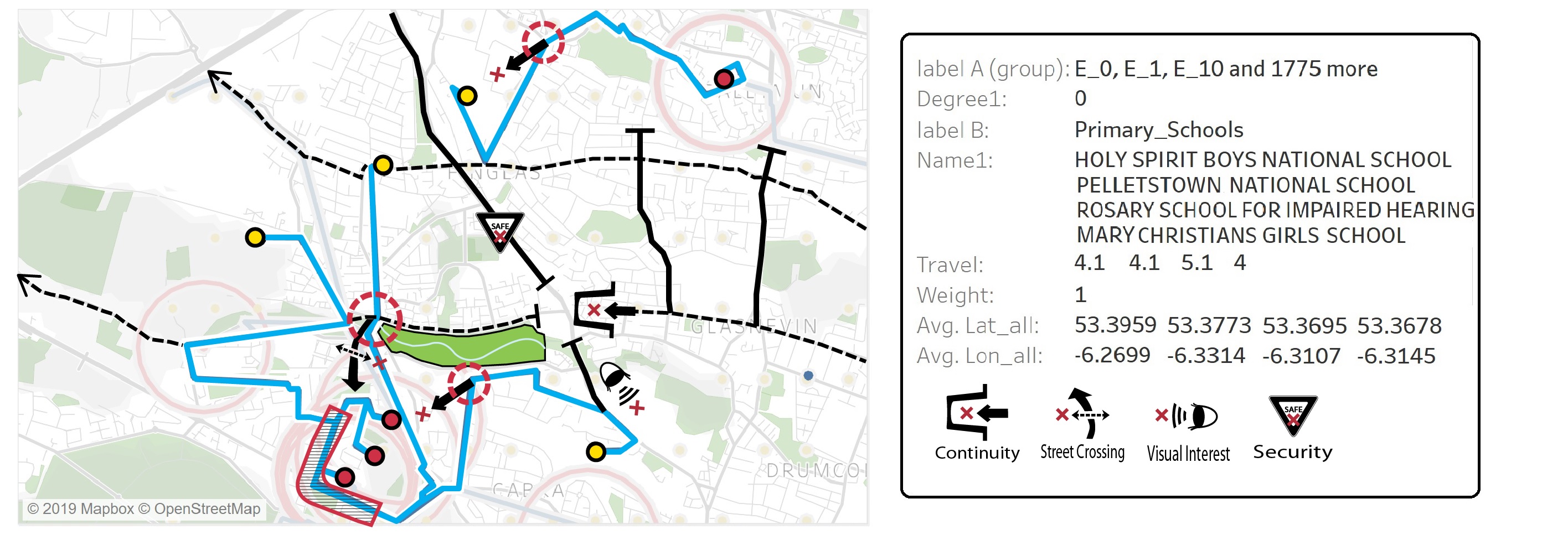}
    \end{center}
    \caption{Site A. HolySpirit School, Pellet School, Rosary School, Mary Christian School and 4 to 5Km of the footpath within the radius of 2Km}
\end{figure}

\noindent
\textbf{B. Footpath from Sports Clubs}: In Figure (15), in spite of its fantastic visual attractions along the seashore, this area fails to provide the shortest, direct routes because of too many gaps and barriers. That is why this area is dominated especially by leisure activities. But it also has several terminal points of the major urban infrastructures such as the M50 ring road, canals, etc. The multiple footpaths are bundled with the main coastal road, performing as a social condenser suggested by \cite{murawski2017introduction}.\\ 
\begin{figure}[!ht] 
    \begin{center} 
        \includegraphics[width=0.9\textwidth]{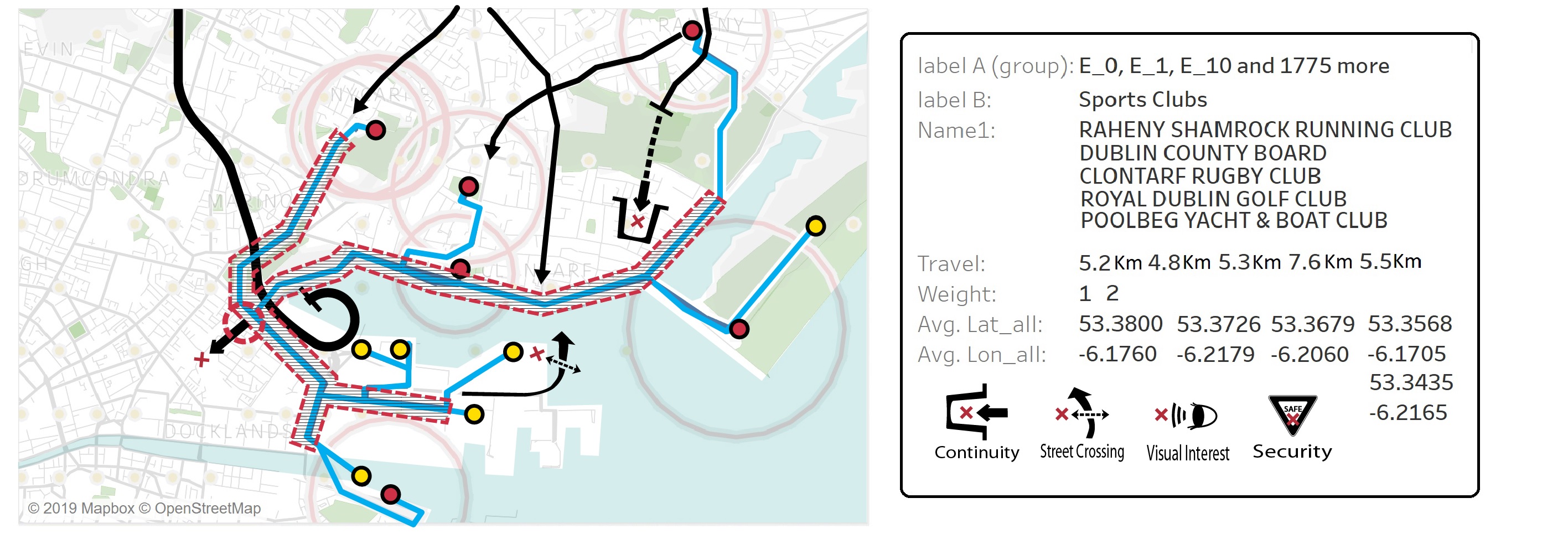}
    \end{center}
    \caption{Site B. Raheny Club, Dublin Board, Rugby Club, Golf Club, boat Club, etc. 5.2 to 7.6Km of the footpath within a radius of 2Km}
\end{figure}

\noindent
\textbf{C. Footpath from Retails (Finglas)}: In Figure (16), the area is a typical urban edge and the pedestrian needs a detour around a huge industrial quarter across the M50 ring road from Finglas. The broken grid divides the area without giving a structure. \\
\begin{figure}[!ht] 
    \begin{center} 
        \includegraphics[width=0.9\textwidth]{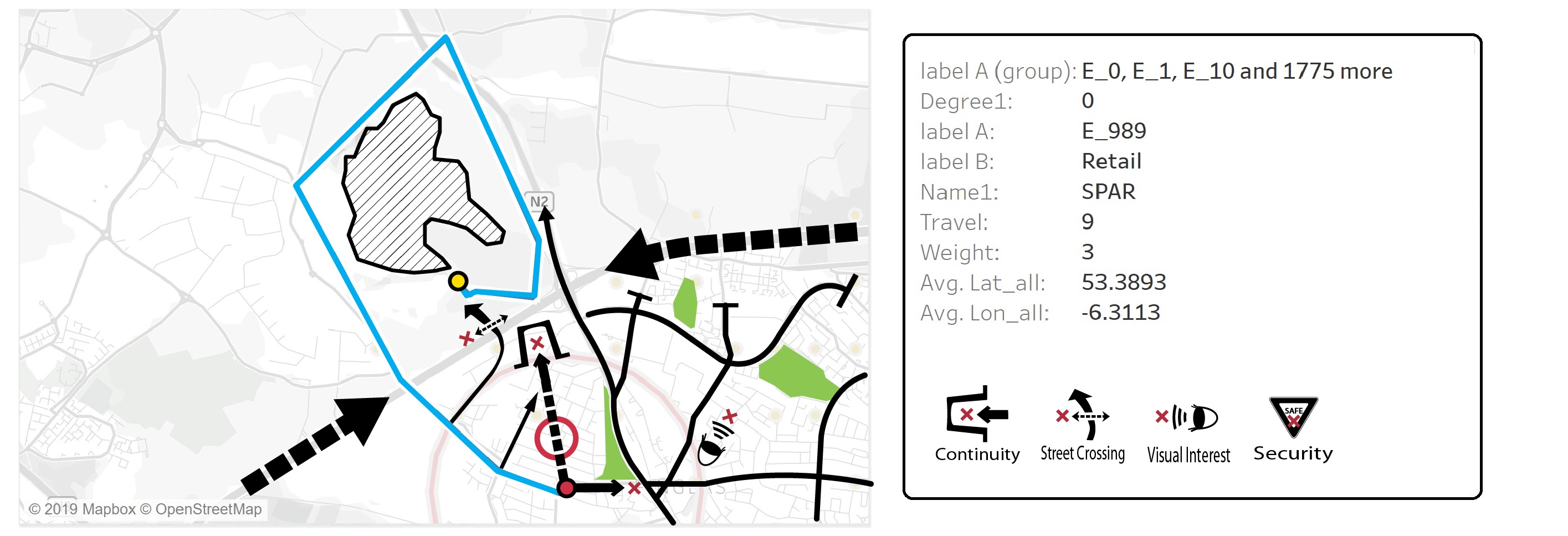}
    \end{center}
    \caption{Site C. Spar convenience store and 9Km of the footpath within the radius of 2Km}
\end{figure}

\noindent
\textbf{D. Footpath from Religious Facilities}: In Figure (17), this area is interlaced with the major ring road M50, offering the main connection to the Dublin International Airport and defining the clear urban edge of Dublin. The origins and the destinations here especially are located on the corners of the grid system where each of the grid cells has a different form and configuration.\\  
\begin{figure}[!ht] 
    \begin{center} 
        \includegraphics[width=0.9\textwidth]{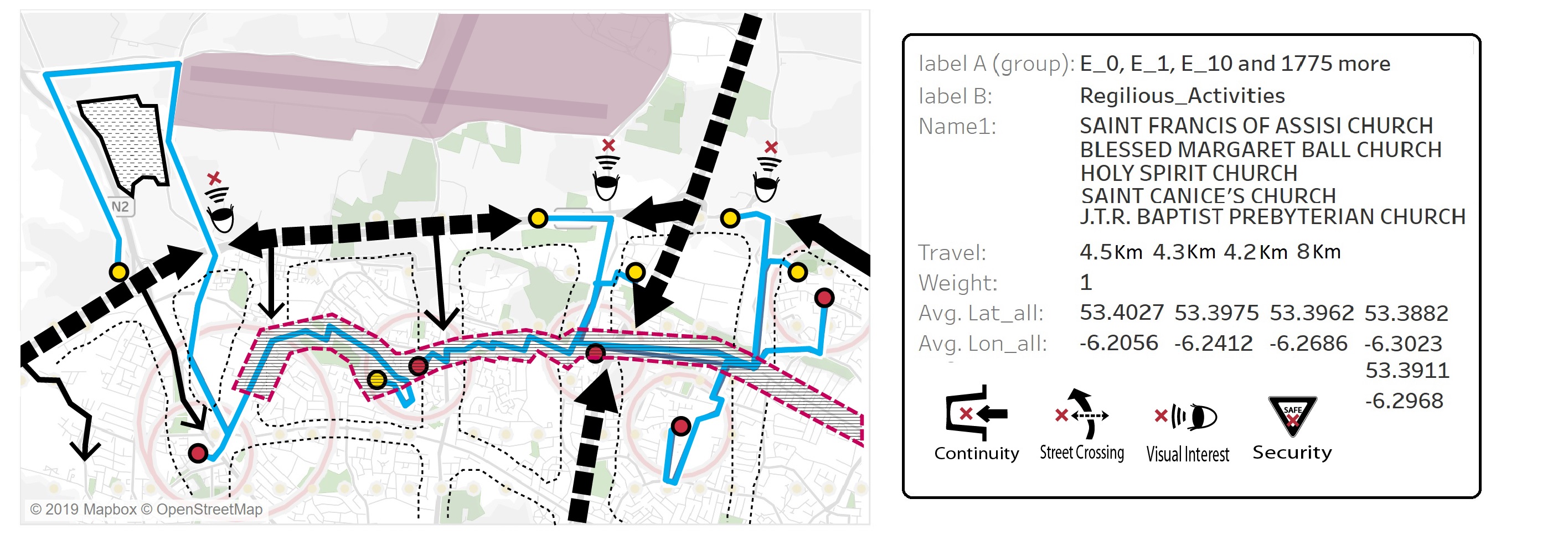}
    \end{center}
    \caption{Site D. Saint Francis Church, MargaretBall Church, Sain Canice Church,J.T.R Church, etc and 4.2 to 8Km of the footpath within the radius of 2Km}
\end{figure}

\noindent
\textbf{E. Footpath from Garda Offices}: In Figure (18), Sitting on the urban edge, this area cannot be characterized as a pedestrian-friendly territory. Due to the expanding nature of this area, the system is not defined and the decent urbanity is not developed yet. \\
\begin{figure}[!ht] 
    \begin{center} 
        \includegraphics[width=0.9\textwidth]{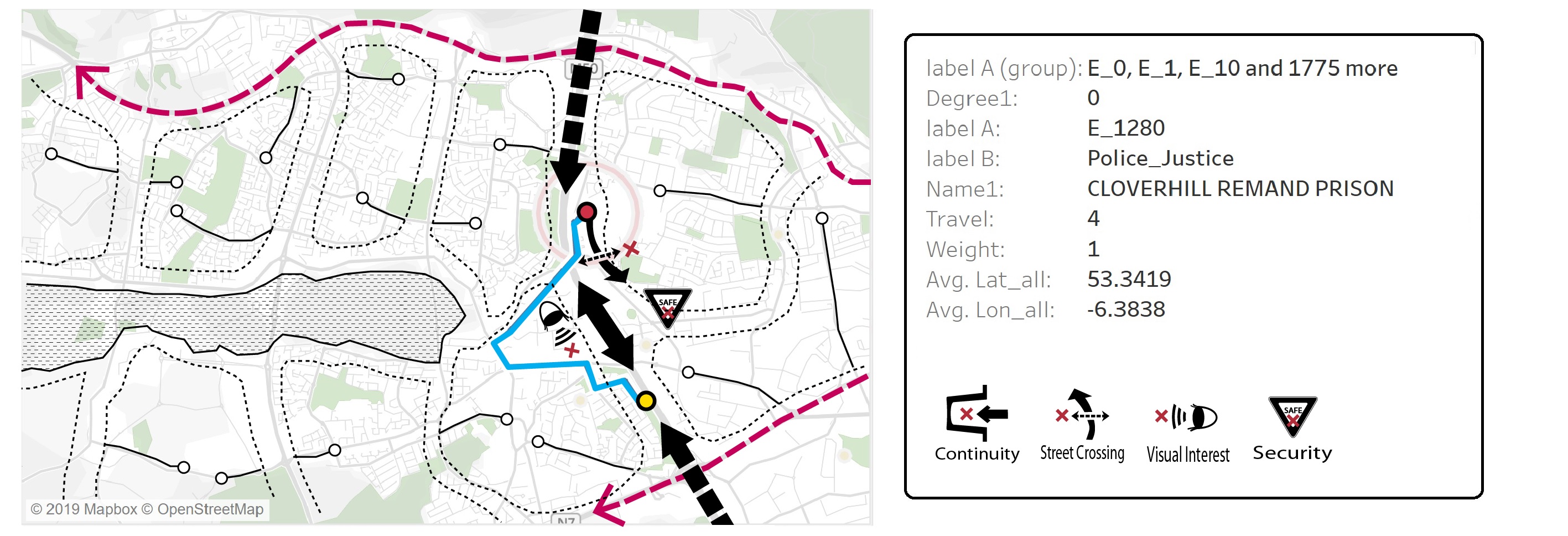}
    \end{center}
    \caption{Site E. Cloverhill Courthouse+Prison and 4Km of the footpath within the radius of 2Km}
\end{figure}

\noindent
\textbf{F. Footpath from Medical Services (Finglas)}: In Figure (19), the amenity point here is located in the heart of Finglas, but the area suffers from a lack of infrastructure. It hardly has any visual attractions and people sometimes feel intimidated and unsafe walking around due to aggressive begging. \\
\begin{figure}[!ht] 
    \begin{center} 
        \includegraphics[width=0.9\textwidth]{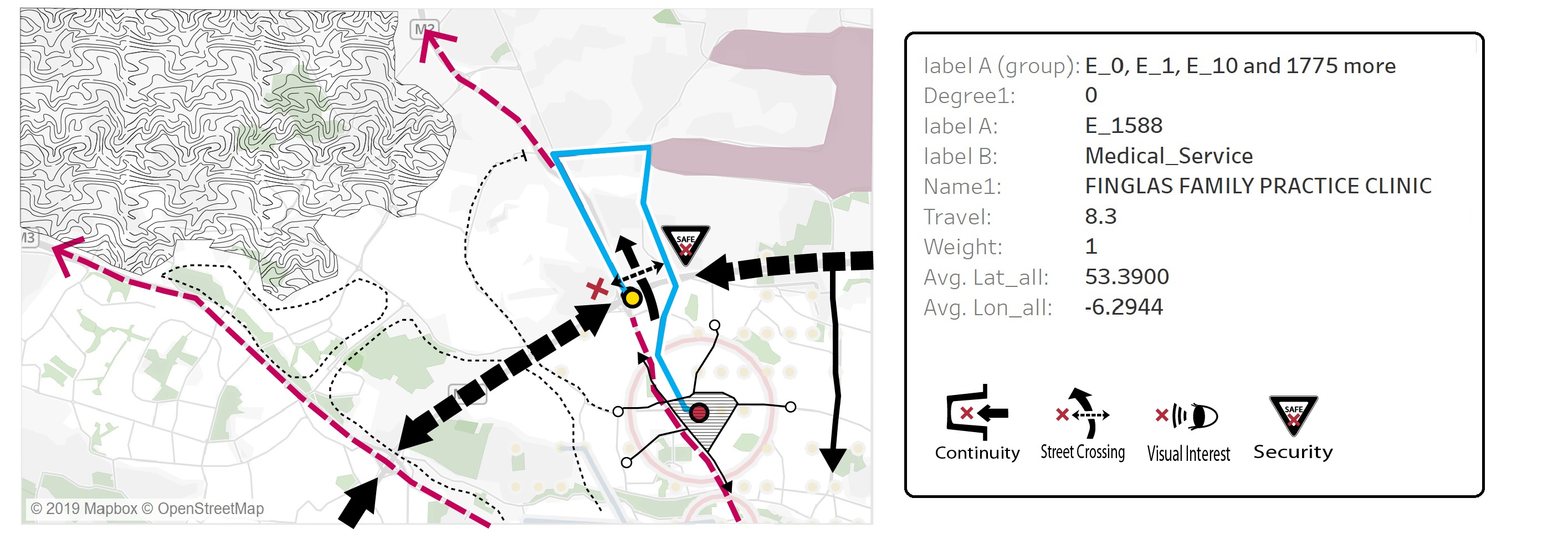}
    \end{center}
    \caption{Site F. Finglas Family Clinic and 8.3Km of the footpath within the radius of 2Km}
\end{figure}

\noindent
\textbf{Summary - Urban Edge and Big Circles:}: In Figure (20), the Urban Edge refers to an urban periphery as a spatial pattern characterized by loosely defined urbanity. Because of its expanding, intermediary nature and its consequent uncertainty, it sometimes suffers from a lack of amenities or awkward spatial definitions. It is not surprising that these big 20 circles indicating the top 20 unwalkable areas are likely to appear around the urban edge of Dublin. It should be noted that these visualizations are best shown interactively and that static screengrabs as used here do not convey the real power and impact of the visualizations.
\begin{figure}[!ht] 
    \begin{center} 
        \includegraphics[width=1\textwidth]{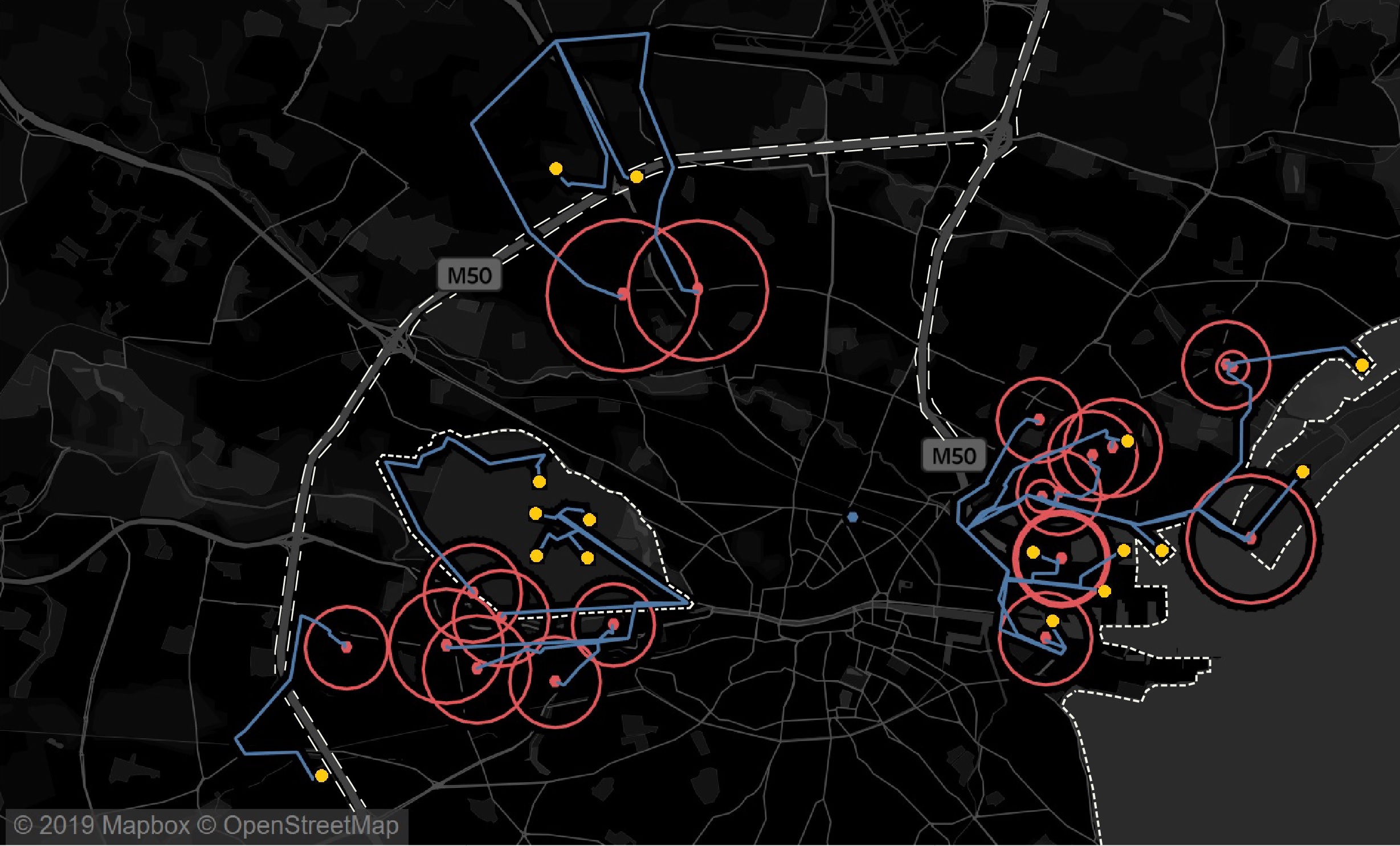}
    \end{center}
    \caption{\textbf{Urban Edge and Big Circles:} Appearing in Ballyfermot, Finglas, Marino, etc. the bigger circles indicate the top 20 unwalkable areas because of the bigger discrepancy. This results from poorly designed relationships between urban infrastructure and inner suburbs. Urbanity around the Urban Edge is too loosely defined.}
\end{figure}

\section{Understanding Walkability: Quantification} We believe that real estate values can be a useful barometer for understanding the impact of walkability and the quality of life in a city. For example, there is a report that in a city in Tennessee, USA, in the early 1980s, (\cite{lerner1999economic}) the quality of life was significantly improved by investing in a walkable environment such as open spaces, parks, trails, etc. This resulted in an increase in annual property tax revenues of 592,000 USD from 1988 to 1996, almost 100 percent higher, because of the increased real-estate values that rose to more than 11 million USD. Such direct monetary return obviously could be the greatest driving force to create a realistic dialogue with urban developers, governors, and individual stakeholders. In this regard, we aim to understand the walkability of Dublin by quantifying its impact on property prices across Dublin City.\\

\noindent
\textbf{Initial Overview of Data:} In our dataset collected from the website: ``\textit{Daft.ie}", the 5362 individual records describe the house prices in 2017 and their locations in the City of Dublin. Given that there is a huge difference in the estimation methods of an appraisal standard between residential and commercial buildings (construction method, capitalization method, etc), we cannot include commercial property prices in our analysis. Instead, we add other extrinsic variables contributing to the change in Dublin house prices. We check the presence of a relationship between housing price and walkability and see if the relationship is linear, additive, quadratic or dependent on other variables in order to quantify the impact of walkability in Dublin. There are, of course, lots of features that affect the formation of the \underline{house prices (continuous)}, but in our study, we pay attention to these 3 predictors particularly:
\begin{itemize}
    \item Type (binary: New/Second-hand)
    \item Distance to the City Centre (continuous: Km)
    \item Difference in the computed travel distances between the Euclidean method and the Google Map method (continuous: Km)
\end{itemize}
\noindent
Our Daft house price dataset already has the ``Type" column. We compute and add the two other variables - ``Unwalkability", ``Distance to the City Center" -, using the same approach that we make previously. \\
\begin{figure*}[!t] 
    \begin{center} 
        \includegraphics[width=\textwidth]{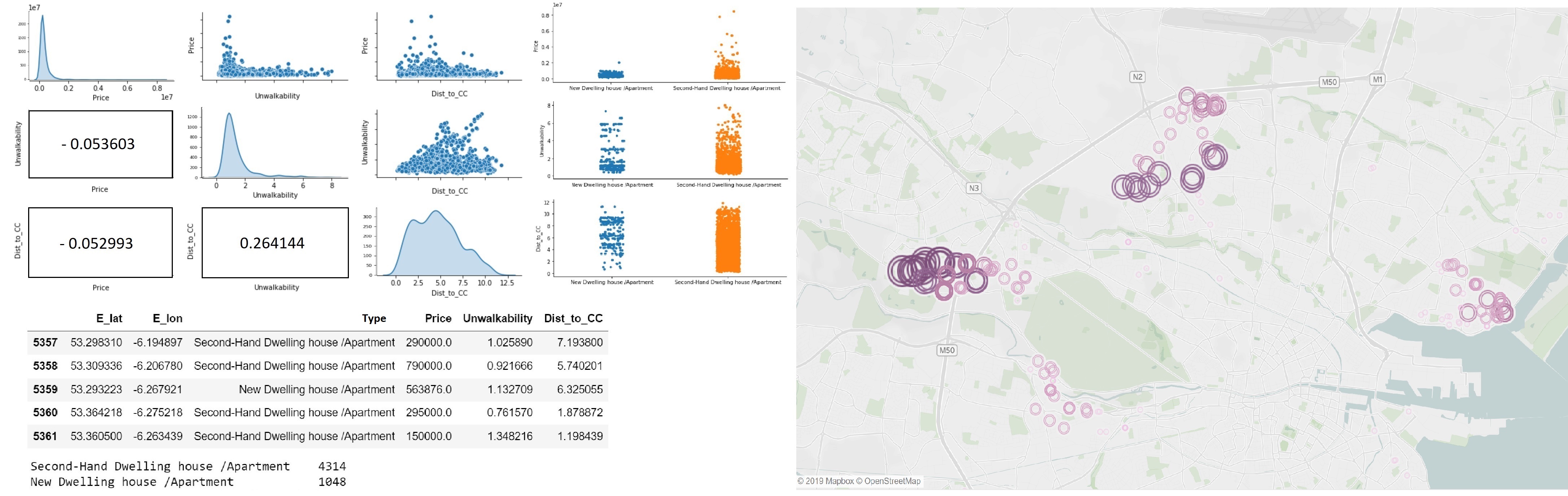}
    \end{center}
    \caption{Matrix of Correlation + distribution + Scatter plot: From the original data, we obtain very small correlation values between variables. Since the distribution of the ``Price" variable is significantly right skewed, we should consider variable transformation method to use the OLS regression without violating any model assumptions.}
\end{figure*}

\begin{figure*}[!b]
  \includegraphics[width=\textwidth]{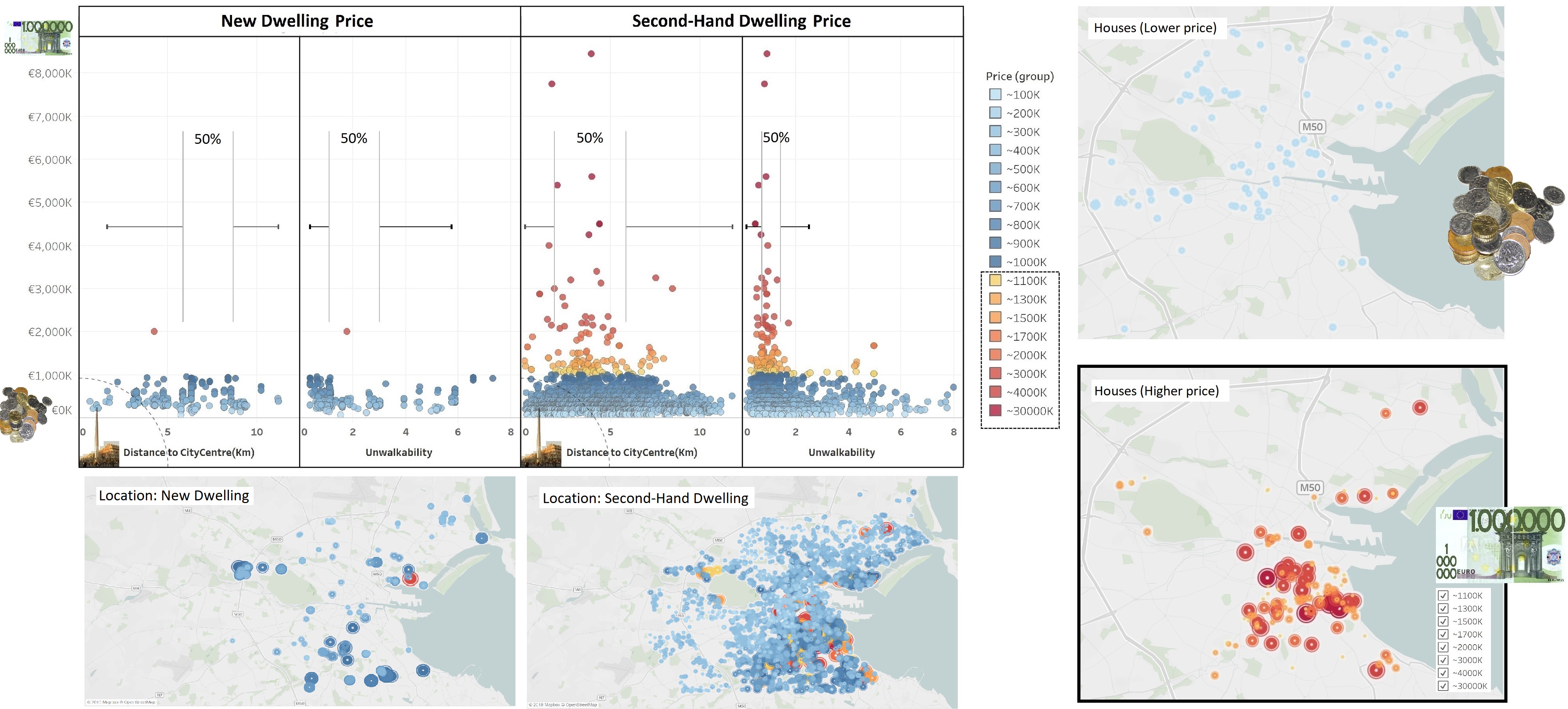}
  \caption{\textbf{Dublin House Property Prices} vs \textbf{Distance to the City Centre} vs \textbf{Walkability:} This series of charts depict the characteristics - location, walkability - of the properties that cost over a million euro. The cheaper properties show greater variance while the expensive properties show smaller variance in their distance to the city centre. As we expect, the expensive properties tend to be closer to the city centre and located in more walkable areas with the un-walkability scores of less than 2.
  Although there is a differentiation between the new property and the second-handed property in the usage variable, the majority of new properties are priced consistently less than 1 million euro and their sample size is too small. Interestingly, the un-walkable areas computed on this house property data appear in the almost same locations as the previous amenity property data.} 
\end{figure*}

\noindent
\textbf{Model Building:} If our goal is to make a prediction as accurately as possible,  we would first have to address all factors such as the number of rooms, materials, technical facilities, historical significance, etc that impact the formation of the property price before we build the model. And we could use the most up-to-date models such as XG-Boost, Bayesian regressor with regularization, etc. However, a simple linear model is useful enough for our study as our primary aim is to quantify the relation between the walkability and the real-estate values. Because of its legibility, a linear model as the most basic machine learning algorithm is often used as a benchmark for more advanced models. Let's say our model equation is given by: \( \vn{House Price}_i = \beta_0 + \beta_1\vn{Usage}_i + \beta_2\vn{Distance to CC}_i + \beta_3\vn{Walkability}_i + \epsilon_i \) where ``Usage" is the only binary variable (New/Second-hand). Now we take the journey for the model development through a series of statistical tests justifying the use of the OLS regression on our data such as 1. Normality and Independence, 2.Non-Multicollinearity, 3. Homoscedasticity, 4. Linearity. \\

\noindent
\textbf{1. Normality and Independence of the Residuals:} As can be seen from the Matrix of Correlation + Distribution + Scatter plot in Figure (21), the data follow more of skewed normal distributions. Given that our sample size is large enough (N= 5362), the model is likely to violate the assumption of normality of the residuals. A common way to correct this would be transforming variables to make the residuals normal. For example, we first take a log transformation of the ``House Price", then see if the distribution fits a better model. We can also consider a log of the ``walakbility" or a square-root of the ``Dist-to-CC", but in our case, they do not make a significant improvement in terms of the R-Squared measure. Hence, we simply choose the Exponential Model with the equation given by: \( \vn{House Price}_i = e^{\beta_0 + \beta_1\vn{Usage}_i + \beta_2\vn{Distance to CC}_i + \beta_3\vn{Walkability}_i + \epsilon_i} \) by taking a log of ``House Price" to meet the condition of normality for further analysis.

In order to test the assumption of the independence in the model residuals, we can use the Durbin-Watson test, looking at the residuals separated by some lag and the test statistics formula is given by: \( DW=\frac{\sum_{i=2}^{N}((y_{i}-\hat{y}_{i})-(y_{i-1}-\hat{y}_{i-1}))^2}{\sum_{i=1}^{N}(y_{i}-\hat{y}_{i})^2} \) where lag is 1 here, and \(y_{i}-\hat{y}_{i}\) refers to the individual residual. The null hypothesis that the residuals of our model are uncorrelated, against the alternative hypothesis that autocorrelation exists. The statistic values range from 0 to 4, and a value around ``2" suggests the failure to reject the null hypothesis. Our first model has the Durbin-Watson statistics value of \textbf{1.91629}; therefore, it is safe to say that its residuals are all independent of each other.\\
\begin{figure}[!ht] 
    \begin{center} 
        \includegraphics[width=0.8\textwidth]{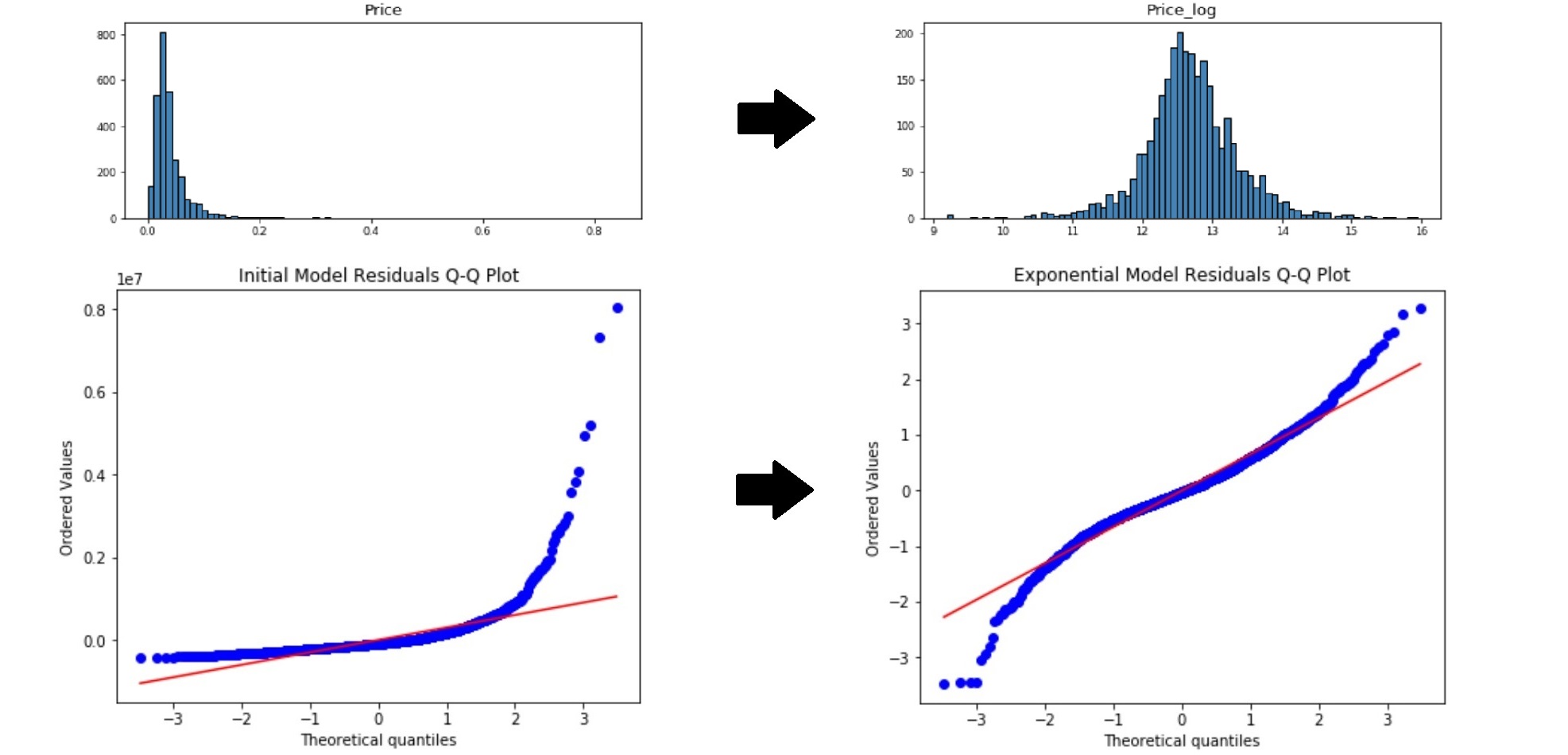}
    \end{center}
    \caption{Top: The distribution of log-transformed dependent variable; Bottom: Model's residual QQ-plot which uses the fractiles of residual distribution versus the fractiles of a normal distribution having the same mean and variance; Left: OLS model's residual plot; Right: Exponetial model's residual plot.}
\end{figure}

\noindent
\textbf{2. Non-Multicollinearity between Predictors:} In Figure (24), we can check for the presence of multicollinearity, using VIF(Variance Inflation Factor) calculated by taking the the ratio of the total variance of model's coefficients divide by the variance of each individual coefficient. Using the formula given by: \(VIF=\frac{1}{1-R^2}\) where \(R^2\) is yielded by the linear relationship between the reference predictor and others, we can inspect the VIF for each predictor. The VIF exceeding 10 is obviously highly problematic. we can say that multicollinearity is present, thus consider L1/L2 regularization. The table below suggests that we do not have to discard any predictors or use a regularization term because of the VIF values smaller than 2.\\
\begin{figure}[!ht] 
    \begin{center} 
        \includegraphics[width=1\textwidth]{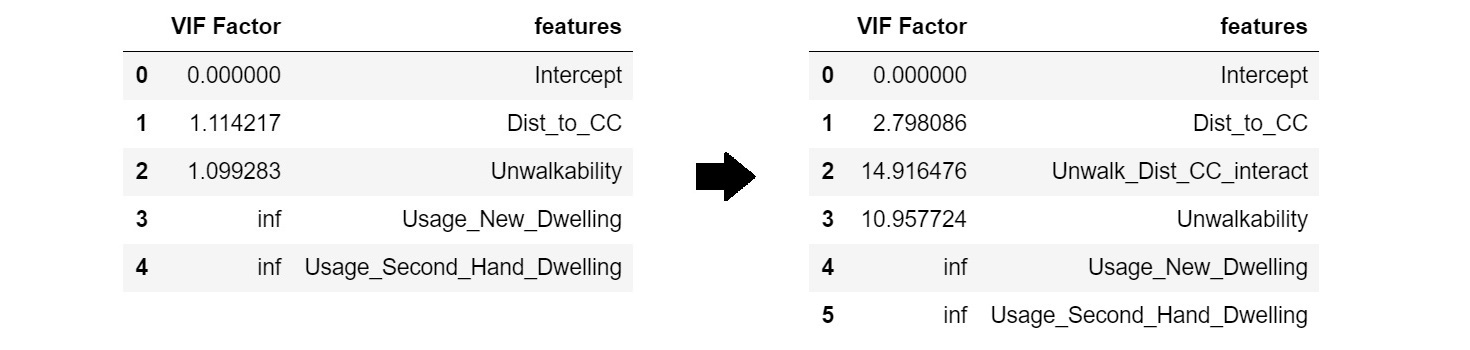}
    \end{center}
    \caption{VIF values: The model predictors does not have a significant VIF. In other words, they do not explain the same variance within this data. VIF is not defined for categorical data.}
\end{figure}

\noindent
\textbf{3. Homoscedasticity of the Residuals:} There are multiple ways to detect the violations of homoscedasticity. One might look at the two scatter plots such as ``residuals vs predicted values" and ``residual vs predictors" to check the consistent variance of the residuals. To be more scientific, we can use statistical test such as NCV, Bruesch-Pagan or White, but the Breusch-Pagan test only checks for the linear form of heteroscedasticity, hence we prefer to use the White test to detect more general form of heteroskedasticity. The test statistics formula is given by: \(F=\frac{(n-2)R_{{\epsilon}^2}^2}{1-R_{{\epsilon}^2}^2}\) or \({\chi}^2=nR_{{\epsilon}^2}^2\) where \(R_{{\epsilon}^2}^2\) refers to the R-squared value from the auxiliary regression: \(\hat{\epsilon}^2_{i}=\delta_{0} + \delta_{1}\hat{y_{i}} + \delta_{2}\hat{y_{i}}^2\) where \(\hat{y_{i}}\) is our fitted values. As stated in the formula, the test statistic approximately follows both F-distribution and Chi-square distribution with the null hypothesis: ``the error variances are all equal". The degrees of freedom for the F-test are ``2" in the numerator and ``n – 3" in the denominator. The degrees of freedom for the chi-squared test are ``2". The test result of our model is 34.73 for W-Statistics with p-value of 0.0000298 and 4.38 for F-Statistics with p-value of 0.0000281, which indicates that our p-value is far less than 0.05, so all two tests agree that our model is heteroscedastic. To correct this violation, one can think of a Box-Cox transformation or WLS (Weighted Least Squares) estimates to stabilize the variance of the residuals. Let's recall that we have an exponential model: \( y_{i} = e^{\beta_0 + \beta_1\vn{Usage}_i + \beta_2\vn{Distance to CC}_i + \beta_3\vn{Walkability}_i + \epsilon_i} \), which estimates parameters: \(\beta_0, \beta_1, \beta_2, \beta_3\) by minimizing the sum of squared residuals given by: $$ SSE=\sum_{i=1}^{N}(y_{i} - e^{\beta_0 + \beta_1\vn{Usage}_i + \beta_2\vn{Distance to CC}_i + \beta_3\vn{Walkability}_i})^2 $$\\ However, if we give our observations some weights, the sum of squared residuals becomes: $$=\sum_{i=1}^{N}W_{i}(y_{i} - e^{\beta_0 + \beta_1\vn{Usage}_i + \beta_2\vn{Distance to CC}_i + \beta_3\vn{Walkability}_i})^2 $$ where \(W_{i} \propto \frac{1}{\sigma_{i}^2}\) then the WLS estimates have smaller standard errors than previous OLS estimates. \(\sigma_{i}^2\) is a variance of the residuals which is \(E[\epsilon_{i}^2]-E[\epsilon_{i}]^2 \), but \(E[\epsilon_{i}]\) is equal to 0, thus \(\sigma_{i}^2\) becomes \(E[\epsilon_{i}^2]\) and that is calculated by \(\frac{\sum_{i}^{N}W_{i}(y_{i}-\hat{y_{i}})^2}{N-p}\)\\
\begin{figure}[!ht] 
    \begin{center} 
        \includegraphics[width=0.8\textwidth]{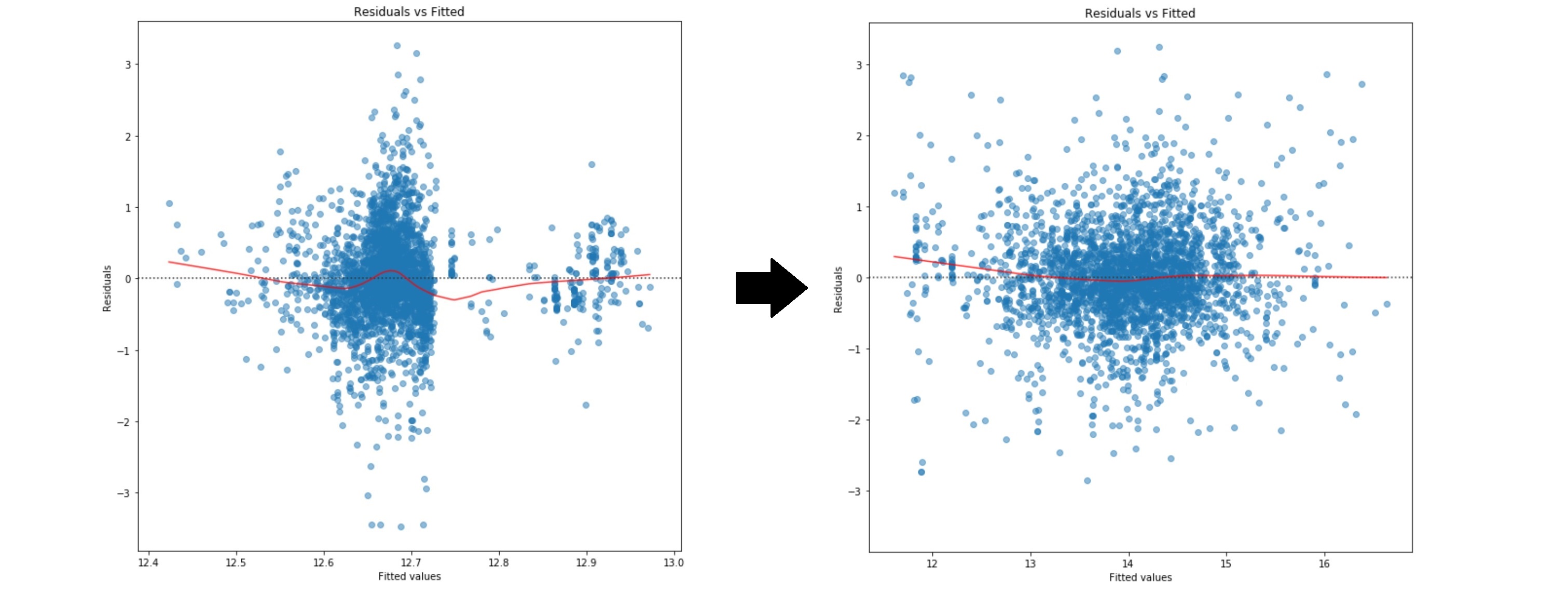}
    \end{center}
    \caption{Residual vs Fitted values plots based on our Exponential Model; Left: before WLS; Right: after WLS}
\end{figure}

\noindent
From the chart in Figure (25), although the magnitude of the residuals' variance becomes smaller, there still appears to be some insignificant heteroscedasticity. It seems that WLS may not always be able to capture all of the heteroscedasticity.\\

\noindent
\textbf{4. Linearity between the Response and Predictors:} Linearity also can be revealed from the Residual versus Fitted value plot above in Figure (26), by the symmetrically distributed data points around the horizontal line. In the case of multiple regression, one can check the systematic patterns in plots of the Residuals versus individual predictors as well. As can bee seen from the chart above, we can say our WLS Exponential model shows a very clear linearity between the response and the predictors. \\
\begin{figure*}[!ht] 
    \begin{center} 
        \includegraphics[width=0.8\textwidth]{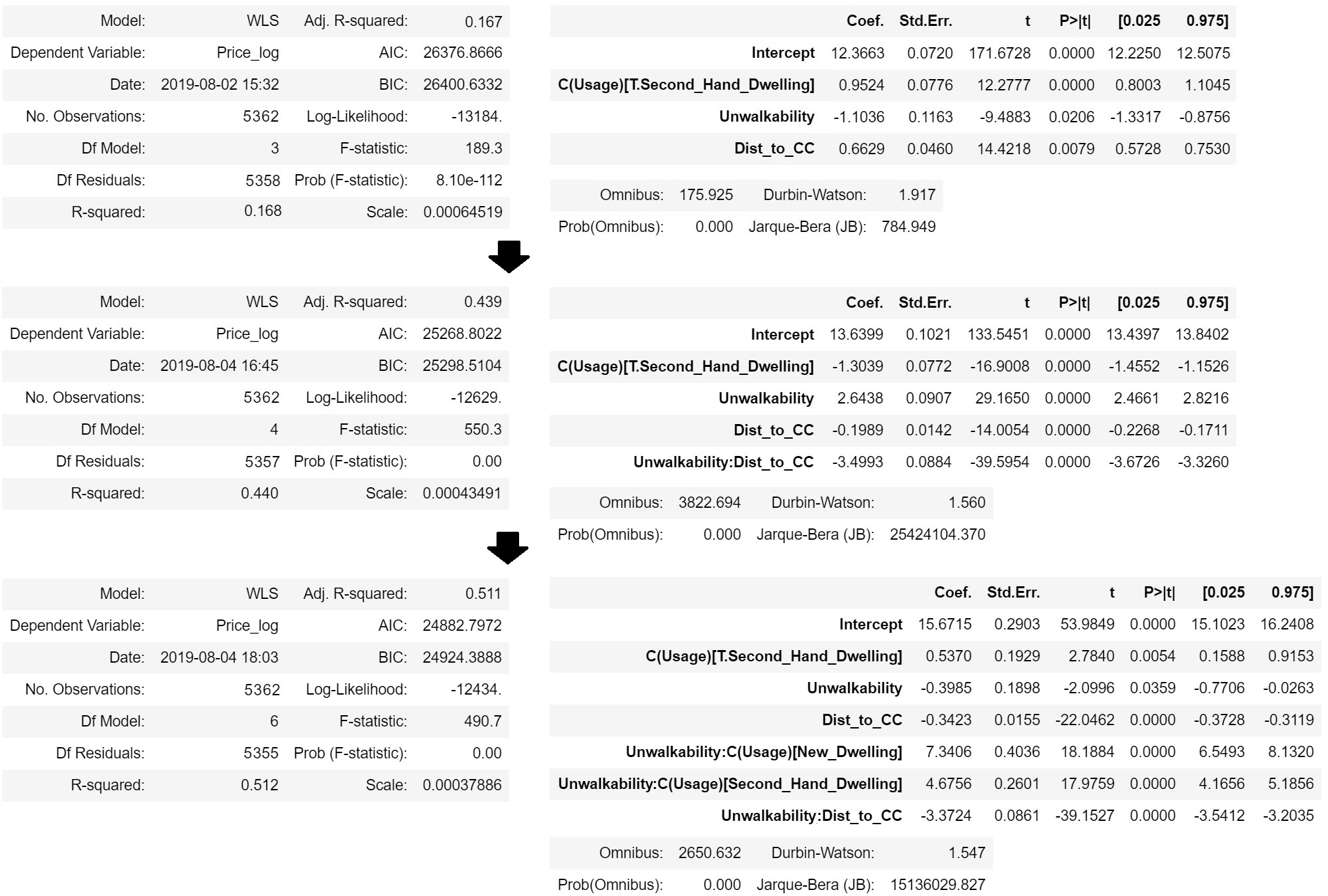}
    \end{center}
    \caption{Introduction of Interaction Terms and WLS Model Output: \(R^2\), F-statistics, Coefficients, Confidence Intervals, P-values, Durbin Watson test. etc}
    \label{EEE-24} 
\end{figure*}

\noindent
\textbf{Results:} We run the exponential model, using the Weighted Least Square technique to rectify the violation of the regression assumptions such as Normality and Homoscedasticity. We add the interaction terms - \(\vn{X_2X_3}\) and \(\vn{X_2X_1}\) - to increase the \(\vn{R^2}\) value.

\begin{itemize}
    \item \textbf{Significant F/T Statistics:} The F-test statistics - 490.7 (\(P<0.05\)) indicate our full model is better than the null model which implies there is a clear relationship between the dependent variable and predictors. The t-test statistics of the predictors also tell us all model coefficients including the two interaction terms are significant.\\
    
    \item \textbf{Low \(R^2\) value:} The low \(\vn{R^2}\) value - 0.5 - in our final model implies we have overlooked other important factors that explain the formation of the Dublin house price. This is not surprising as the 3 predictors we use - \(\vn{X_1}:\)``Dwelling type", \(\vn{X_2}:\)``Unwalkability", \(\vn{X_3}:\)``Distance to the city centre",  - are obviously not enough to describe the full mechanism of the real estate market. However, given that our main interest in particular is to quantify the contribution of the Unwalkability variable \(X_2\) to the house price increase, it is still meaningful to track down the improvement of the \(R^2\) value as we add or remove some additional predictors such as an interaction term generated by the \(X_2\). \\
    \item \textbf{Two Interaction terms:} Through out the model building process, the best \(R^2\) value is obtained when we add the interaction terms involving the ``Unwalkability" - \(\vn{X_2X_3}\) and \(\vn{X_2X_1}\). This implies the unique effect of ``Unwalkability" on ``House price" is not limited to \(\beta_2\vn{X_2}\) but also relies on the values of \(\beta_3\vn{X_3}\) and \(\beta_1\vn{X_1}\). This is the ``simultaneous influence of two variables" on a third. Hence, our interpretation of the model coefficients drastically changes as we add the interaction terms due to multiplicative relations between variables. For example, we should see if the relationship between the ``Unwalkability" and ``house price" is different by ``Type" (New / second hand), or by variance in ``Distance to City Centre". From the model output in Figure (26), looking at the coefficients of the first interaction terms (1): \(+4.67\vn{X_2X_1}\) when \(\vn{X_1} = ``SecondHand"\), and \(+7.34\vn{X_2X_1}\) when \(\vn{X_1} = ``New"\), one can see that houses in more unwalkable area tend to be pricier in both ``new" and ``second hand" groups but the relationship is much more dramatic in the ``new" group than in the ``secondhand". The second interaction term (2) has the coefficients: \(-3.37\vn{X_2X_3}\), thus it is safe to say that houses in unwalkable area with a large distance from city centre tend to be cheaper because of its negative coefficient. 
\end{itemize}

\begin{figure*}[!b]
  \includegraphics[width=\textwidth]{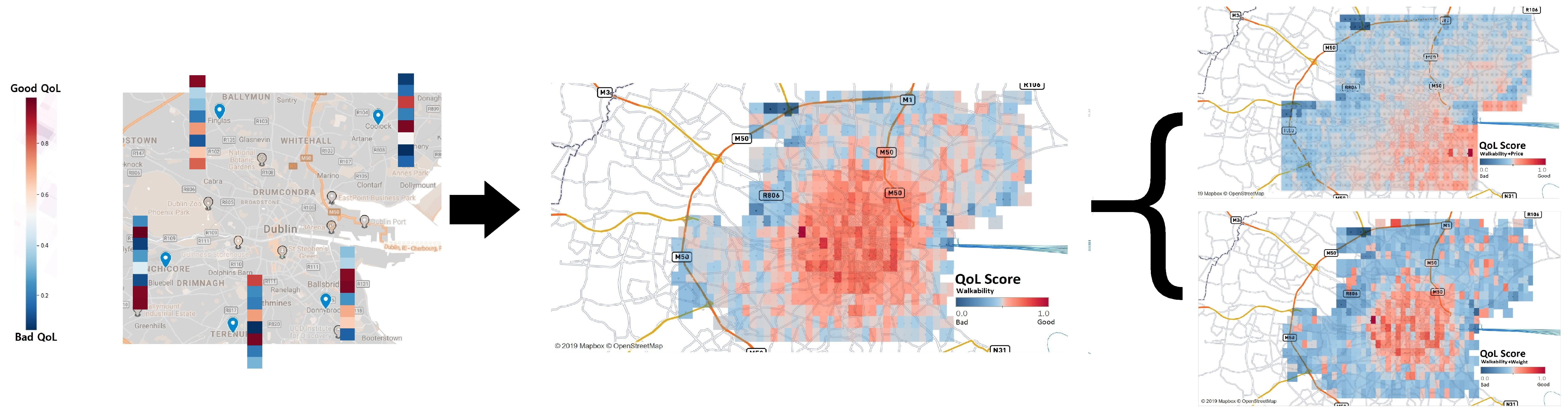}
  \caption{\textbf{Dublin QoL matrix}: We expand the scope of QoL analysis from the walkability perspective to social activities (importance weights) and house prices. It tells us the QoL in the unwalkable areas can get worse due to lack of social activities. Although the QoL map in terms of house prices manifests interesting spectacle where the QoL score goes up as approaching to Dublin Bay, the house price factor cannot be definitive and called for more research.} 
\end{figure*}

\section{Conclusion} This study suggests how to visualise Unwalkable areas in Dublin and how to understand the cost of unwalkability from an economic point of view. From the Tableau visualisation, we better detect the unwalkable areas across the whole Dublin city by mapping the pedestrian footpaths and their travel distances to the points of interest. We found that the majority of unwalkable areas are likely to appear around the urban edge of Dublin. To quantify the effect of unwalkability, we set up the model as below: 
\begin{multline*}
\log(HousePrice_{i}) = 15.67 + 0.54*\vn{Type}_i \\
- 0.39*\vn{Unwalkability}_i \\
- 0.34*\vn{DistCityCentre}_i \\ + 7.34*(\vn{Unwalkability}_i*\vn{Type 1}_i) \\
+ 4.67*(\vn{Unwalkability}_i*\vn{Type 0}_i) \\
- 3.37*(\vn{Unwalkability}_i*\vn{DistCityCentre}_i)
\end{multline*}
The coefficient can be interpreted as the effect in unit change in terms of the House Price. In Dublin, for every increase in the ``Unwalkability", there will be \\
\(e^{15.67 + 0.54 - 0.39 + 7.34 - 3.37*DistCityCentre}\) change in the ``house price" when the property type is ``new", or there will be \\
\(e^{15.67 - 0.39 + 4.76 - 3.37*DistCityCentre}\) change in the ``house price" when the property type is ``second hand". Of course we acknowledge that our model is far from perfect as there are many other important variables affecting the mechanism of calculating Dublin housing prices. However, our study exemplifies how the concept of walkability as a spatial quality can be capitalised by the house price market, which is important for politicians, regional councils, or other stakeholders who have to care for these regions.   

\subsection*{Acknowledgments}
We thank Dublin City Council for the provision of data used in this work and for feedback on the concepts behind this.

\bibliographystyle{chicago}

\bibliography{Bibliography-MM-MC}

\begin{thebibliography}{}

\bibitem[\protect\citeauthoryear{Associates}{Associates}{2003}]{MeasuringWalkability}
Associates, L. (2003).
\newblock “kansas city walkability plan”.
\newblock {\em Kansas City Council Archive, Missouri\/}~{\em 11\/}(030211), 15.

\bibitem[\protect\citeauthoryear{d'Acci}{d'Acci}{2019}]{d2019quality}
d'Acci, L. (2019).
\newblock Quality of urban area, distance from city centre, and housing value. case study on real estate values in turin.
\newblock {\em Cities\/}~{\em 91}, 71--92.

\bibitem[\protect\citeauthoryear{Glaeser}{Glaeser}{2011}]{glaeser2011triumph}
Glaeser, E. (2011).
\newblock {\em Triumph of the city: How urban spaces make us human}.
\newblock Pan Macmillan.

\bibitem[\protect\citeauthoryear{Lerner, Poole, et~al.}{Lerner et~al.}{1999}]{lerner1999economic}
Lerner, S., W.~Poole, et~al. (1999).
\newblock The economic benefits of parks and open space: How land conservation helps communities grow smart and protect the bottom line.

\bibitem[\protect\citeauthoryear{Li, Santi, Courtney, Verma, and Ratti}{Li et~al.}{2018}]{li2018investigating}
Li, X., P.~Santi, T.~K. Courtney, S.~K. Verma, and C.~Ratti (2018).
\newblock Investigating the association between streetscapes and human walking activities using google street view and human trajectory data.
\newblock {\em Transactions in GIS\/}~{\em 22\/}(4), 1029--1044.

\bibitem[\protect\citeauthoryear{Murawski}{Murawski}{2017}]{murawski2017introduction}
Murawski, M. (2017).
\newblock Introduction: crystallising the social condenser.
\newblock {\em The Journal of Architecture\/}~{\em 22\/}(3), 372--386.

\bibitem[\protect\citeauthoryear{Rafiemanzelat, Emadi, and Kamali}{Rafiemanzelat et~al.}{2017}]{rafiemanzelat2017city}
Rafiemanzelat, R., M.~I. Emadi, and A.~J. Kamali (2017).
\newblock City sustainability: the influence of walkability on built environments.
\newblock {\em Transportation research procedia\/}~{\em 24}, 97--104.

\bibitem[\protect\citeauthoryear{Sinnott}{Sinnott}{1984}]{EuclideanDistance}
Sinnott, R. (1984).
\newblock “virtues of the haversine”.
\newblock {\em Sky and Telescope\/}~{\em 68\/}(2), 159.

\bibitem[\protect\citeauthoryear{Xu, Belyi, Bojic, and Ratti}{Xu et~al.}{2018}]{xu2018human}
Xu, Y., A.~Belyi, I.~Bojic, and C.~Ratti (2018).
\newblock Human mobility and socioeconomic status: Analysis of singapore and boston.
\newblock {\em Computers, Environment and Urban Systems\/}~{\em 72}, 51--67.

\end{thebibliography}
\end{document}